\documentclass[aps,pre,showpacs,twocolumn]{revtex4-1}
\usepackage[utf8x]{inputenc}
\usepackage{verbatim}
\usepackage{graphicx}
\usepackage{subfigure}
\usepackage{color}
\usepackage[]{amsmath}
\usepackage{amsfonts}
\usepackage{amssymb}
\usepackage{soul}
\begin{document}
\title{Multi-scale modeling of dislocation-precipitate interactions in Fe:\\ 
from molecular dynamics to discrete dislocations}

\author{Arttu Lehtinen$^1$, Fredric Granberg$^2$, Lasse Laurson$^1$, Kai Nordlund$^2$, and Mikko J. Alava$^1$}
\affiliation{$^1$Department of Applied Physics, Aalto University School of Science,
P.O. Box 11100, FIN-00076 Aalto, Espoo, Finland}
\affiliation{$^2$Department of Physics, P.O. Box 43, FIN-00014 University of Helsinki, Finland}
\email{arttu.lehtinen@aalto.fi}


\date{\today}

\begin{abstract}

The stress-driven motion of dislocations in crystalline solids, and thus the ensuing plastic deformation
process, is greatly influenced by the presence or absence of various point-like defects such as precipitates
or solute atoms. These defects act as obstacles for dislocation motion and hence affect the mechanical
properties of the material. Here we combine molecular dynamics studies with three-dimensional discrete 
dislocation dynamics simulations in order to model the interaction between different kinds of 
precipitates and a  $\frac{1}{2}\langle 1 1 1\rangle$ $\{1 1 0\}$ edge dislocation in BCC iron. We have 
implemented immobile spherical precipitates into the ParaDis discrete dislocation dynamics code, with the 
dislocations interacting with the precipitates via a Gaussian potential, generating a normal force acting on 
the dislocation segments. The parameters used in the discrete dislocation dynamics simulations for the 
precipitate potential, the dislocation mobility, shear modulus and dislocation core energy are obtained 
from molecular dynamics simulations. We compare the critical stresses needed to unpin the dislocation 
from the precipitate in molecular dynamics and discrete dislocation dynamics simulations in order to fit 
the two methods together, and discuss the variety of the relevant pinning/depinning mechanisms.      

\end{abstract}

\maketitle

\section{Introduction}

The crucial role of dislocations and their stress-driven dynamics on the mechanical properties 
of metals is a well-established fact. Nevertheless, the underlying mechanisms of how dislocations interact 
with various obstacles such as precipitates, solute atoms or grain boundaries has only recently been 
considered in the context of numerical simulations~\cite{terentyev2008interaction,Koning2003modeling}. Recent developments in computational 
physics have made it possible to study these phenomena on a multiscale level~\cite{Rubia2000multiscale,wirth2004multiscale,dewald2007multiscale,kumar2012multiscale}. Using some approximations, 
massively parallel computers are now capable of performing molecular dynamics (MD) simulations of 
multi-million atom systems for long enough time scales such that integrating the MD results with various 
higher-level, coarse-grained descriptions of the problem at hand becomes meaningful. A relevant example 
of such a coarse-grained description is given by discrete dislocation dynamics
(DDD) simulations, with flexible dislocation lines interacting via long-range stress fields as the 
basic degrees of freedom instead of explicit consideration of the atoms in the crystal lattice~\cite{bulatov2006computer}. Here, 
the basic idea of multiscale modeling is to obtain a set of key parameters from MD simulations, 
and then use the obtained parameter values in DDD simulations, to realistically model a large system with multiple dislocations~\cite{groh2009multiscale}.

Steels are some of the most widely used structural materials in various fields of engineering, due to 
their good properties and versatility. They tend to have a very complex nanostructure, affecting 
the movement of dislocations, and therefore the mechanical properties of the material. One key feature 
of these nanostructures are precipitates of various kinds, either naturally occuring solutes, for instance 
carbides, or man-made particles as in oxide dispersion-strengthened (ODS) alloys~\cite{odette2008recent,hirata2011}. To predict the 
effects of these precipitates on the mechanical properties of a steel sample, knowledge of both the 
atomic scale properties of a single dislocation interacting with different kinds of obstacles, as well 
as those due to the synergetic effect of multiple dislocations interacting with multiple obstacles of 
different kinds, is necessary. Thus, a multiscale modeling framework integrating MD and DDD is needed.

In this work we utilize MD to investigate the atomic scale interaction of an edge dislocation with 
different kinds of obstacles in BCC iron, thus obtaining the parameters necessary to run accurate DDD 
simulations on a larger scale. In order to specifically study precipitate hardening, we have implemented 
a new obstacle datastructure within the DDD code ParaDis~\cite{arsenlis2007enabling}. These 
obstacles are immobile, destructible and they interact with dislocations via a simple Gaussian 
potential. In the literature the usual way to model precipitates is to make them impenetrable obstacles or 
that they impose a constant drag force on the dislocation segments in contact with them~\cite{mohles2004critical,monnet2011orowan}. The Gaussian potential makes it possible for the dislocation to continuously 
penetrate into the obstacle, and we can control the strength of this penetration by tuning the 
precipitate strength parameter. This leads to a physically more accurate description of the 
dislocation-precipitate interaction. 
 
Parameters for the potential are obtained by comparing the dislocation-precipitate 
unpinning stresses of the MD simulations to those of the DDD. Also other relevant parameters such
as dislocation mobility, shear modulus and dislocation core energy are estimated from the MD
simulations, and used in the coarse-grained DDD simulations. The obstacle datastructure is build in general manner so that it is relatively simple to study other dislocation-precipitate interaction potentials in the future.

This paper is organized as follows. First in Sections IIA and IIB we describe in a general fashion 
both the MD and the DDD simulation methods we have used. In Section IIC the precipitate implementation 
for ParaDis is described in detail and in section IID we introduce our multiscale framework for precipitate pinning. Then we present the specific results for the 
dislocation-precipitate interaction of both methods in Sections IIIA and IIIB, respectively. In 
Section IIIC we describe how they are fitted together and compare the unpinning stress, $\sigma_\text{c}$, obtained from 
our simulations to the analytic result of Bacon {\it et al.} \citep{bacon1973effect}. Finally in Section IV we discuss our results and  present our conclusions.

\section{Methods}

In this study, two different computational techniques are utilized, molecular dynamics and 
discrete dislocation dynamics simulations, to be able to investigate dislocation movement 
on a multiscale level. MD simulations are used to extract various parameters (obstacle strength,
shear modulus, dislocation mobility and core energy) which are then used in the DDD simulations.
The results of the two methods are first compared using the same setup in both methods, to fine-tune
the DDD model. Then, DDD simulations may be used to perform realistic studies for much larger 
length and longer time scales.

\subsection{Molecular dynamics}

A classical molecular dynamics code, PARCAS~\cite{nor97,gha97}, was used with a Tersoff like bond 
order interatomic potential, H13, by Henriksson \textit{et al.} for describing FeCrC ~\cite{hen13}. To investigate the 
strength of different precipitates and to estimate the dislocation mobility, a simulation technique 
by Osetsky and Bacon was used~\cite{ose03}.

\begin{figure}
\begin{centering}
\includegraphics[scale=0.08,clip=true,trim= 0 0 0 0]{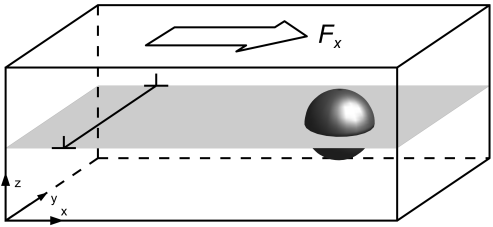}
\caption{(Color online). Schematic diagram of the MD simulation box. The dislocation is positioned on the left side and the precipitate to the right. Shear is generated to the crystal by moving the upper layer of atoms in the box. Under this shear the dislocation moves on its slip plane towards the precipitate.}
\label{fig1}
\end{centering}
\end{figure}

The simulation setup according to~\cite{ose03} 
can be seen in Fig.~\ref{fig1}, where the $x$\st{-}, $y$\st{-} and $z$\st{-}axes are oriented along the $[111]$, 
$[\bar{1}\bar{1}2]$ and $[1\bar{1}0]$ directions, respectively. The uppermost and lowermost 
layers of atoms in $z$-direction are fixed, and the uppermost layers are displaced, relative to 
the lowermost layers, with a constant strain rate to achieve a glide force acting on the dislocation. 
The shear stress induced on the simulation cell can be calculated from $\tau = F_{x}/A_{xy}$, 
where $F_x$ is the total force on the atoms in the fixed block in $x$-direction and $A_{xy}$ 
the area in the $x-y$ plane. The atoms between the two fixed layers were able to move according 
to the Newtonian equations of motion, and a few layers of atoms above the fixed atoms at the bottom 
were also thermally controlled by a Berendsen type thermostat~\cite{ber84}. The same 
method and simulation cell have been previously used in similar investigations in 
Refs.~\cite{gra1,gra2,gra3}. Some of the results are taken from these references and used 
as parameters in the DDD simulations. 

The simulation block for the obstacle simulations was $101 \times 3$, $30 \times 6$ and 
$30 \times 2$ atomic planes, resulting in a cell with the volume $25 \times 21 \times 12$ 
$\mathrm{nm^3}$. Periodic boundary conditions (PBC) were used in both $x-$ and $y-$directions, 
resulting in a length of $21.2\,\text{nm}-d_{p}$ between the obstacles, with $d_{p}$ the 
diameter of the obstacle. This procedure effectively results in an infinite array of obstacles 
of the same size. The simulation method and size were chosen to be comparable with previous results and the choice
will only induce a maximum distortion of 0.5$\,\%$ \cite{ose03}, which is negligible.

The constant strain rate $\gamma$ was $5 \times 10^{7}$ 1/s, resulting in a 
dislocation velocity of 50 m/s, according to the Orowan relation $\gamma = 
b \rho v$, with $b$ the Burgers vector, $\rho$ the dislocation density and $v$ the dislocation 
velocity. To visualize the dislocation core, in order to for instance estimate the dislocation 
mobility, we used the program OVITO and the adaptive common neighbor analysis implemented in 
the program~\cite{stu10}.  

To obtain the elastic parameters of iron to be used in the DDD simulations, we used a 
smaller simulation cell of about 10000 atoms. The parameters of interest were the Poisson ratio 
and the shear modulus. To obtain the Poisson ratio we elongated the box in one direction and 
calculated shrinkage in the other directions, and from that calculated the ratio. To 
obtain the shear modulus we fixed the bottom layer of atoms and the uppermost layer of atoms, 
and shifted the upper layers inducing a shear on the box. The box was then relaxed at a 
temperature $T=750 \,\text{K}$, followed by the calculation of the virial, and subsequently that of the 
shear modulus in the  $[111]$ direction.
 
 To investigate the effect of the distance between the obstacles, 
we used the same simulation cell as described in the previous paragraph, 
but varied the amount of atomic planes in the $y$-direction to obtain 
different lengths between the obstacles in the infinite array over the 
periodic boundaries. The length dependence is crucial to know to be able 
to get the right unpinning stresses for obstacles that are separated by 
another distances than the one(s) studied in MD, discussed in more detail 
in Section III C. In the investigation of the length dependence we used 
spherically fixed atoms, with the diameter $2$ nm, as an obstacle. The 
obstacles we studied in MD to get the qualitive results were spherical 
cementite ($Fe_{3}$C) precipitates, of the sizes $1\,\text{nm}$, $2\,\text{nm}$  
and $4\,\text{nm}$ [21]. The cementite precipitates had the orthorhombic 
lattice structure according to the space group Pnma (no. $62$). The 
precipitates contained $18$, $116$ and $940$ carbon atoms for the sizes 
$1\,\text{nm}$ , $2\,\text{nm}$  and $4\,\text{nm}$, respectively. The 
precipitates were cut out from a pristine block of cementite to the right 
size, compressed by $5~\%$ and placed inside a void in the block with 
the dislocation. The block with the precipitate was then relaxed before 
the straining of the cell started. The potential showed a flow stress at 
the used strain rate and temperature. The flow stress has been subtracted 
from the obtained value for unpinning stress, to determine the absolute 
strength of only the obstacle. To investigate the velocity of the edge 
dislocation, a pristine block of BCC Fe with an edge dislocation was 
used. The dimensions were the same in the $y$- and $z$-directions as in 
the previous paragraph, but the $x$-direction was only $60 \times 3$ 
atomic planes. We used 6 different forces to shear the block, by fixing 
the atoms at the bottom and applying the force to the few uppermost layers 
of atoms, and thereby applying a constant glide force on the dislocation.

\subsection{Discrete dislocation dynamics}

In DDD simulations, the relevant degrees of freedom are flexible dislocation lines, consisting
of discrete segments. The dislocation lines move and change shape when subject to stresses. 
The total stress acting on a dislocation segment consists of the 
external part, resulting from the deformation of the whole crystal, and of the internal, 
anisotropic stress-fields generated by the other dislocation segments within the crystal. 
The latter stress fields are computed using the well-known results of linear elasticity 
theory. Near the dislocation core, local interactions, such as junction formation, annihilation,  
etc., are introduced phenomenologically using smaller scale simulation methods (e.g. MD) and 
experimental results as guidelines. The strength of the long-range stress field of dislocations  
decays as $\sim \frac{1}{r}$ with distance $r$, leading to $O(N^2)$ computational cost, if calculated directly. This, together with the fact that the topology of the dislocation lines changes in time, makes parallel simulation algorithms a necessity.

There are many existing DDD codes. We have chosen ParaDis ~\cite{arsenlis2007enabling} because of its 
good documentation, parallel scalability and clear modular structure. In ParaDis dislocations 
are modeled using a nodal discretization scheme: dislocation lines are 
represented by nodal points connected to their neighbors by dislocation segments. Forces 
between segments of nearby nodes and self-interaction of dislocations are calculated with 
explicit line integrals. Far-field forces are calculated from the coarse grained dislocation 
structure using a multipole expansion. In real materials, the motion of dislocations are 
subject to constraints which depend on the underlying crystal structure and the nature of 
the dislocations (e.g. screw or edge) in a complicated manner. These details are encoded 
in the material-specific mobility function which relates the total stresses experienced by 
dislocations to their velocities.

\subsection{Implementation of precipitates in DDD}

There are no point-like arbitrarily strong pinning defects implemented in the default 
version of ParaDis. Only objects which have a dislocation nature, e.g. lines and loops, 
are readily implemented. In order to remedy this, we have added a new precipitate datastructure 
into the ParaDis code. These precipitates are spherical and immobile and they generate a 
Gaussian potential $U(r)=Ae^{-\frac{r^2}{R^2}}$ around them. Thus, the interaction force between 
dislocation and precipitate is
\begin{align}
\label{eq2}
F(r)&=-\nabla U(r) =\dfrac{2 A r e^{-\frac{r^2}{R^2}}}{R^2 }\,, \nonumber\\
\end{align}
where $r$ is the distance to the center of the precipitate, $R$ is the radius of the 
precipitate, and $A$ is a parameter quantifying the pinning strength of the precipitate.
The Gaussian potential was chosen for simplicity: it describes a short-range interaction 
with a continuous force-field.

\begin{figure}
\begin{centering}
\includegraphics[scale=0.3,clip=true,trim= 0 0 0 0]{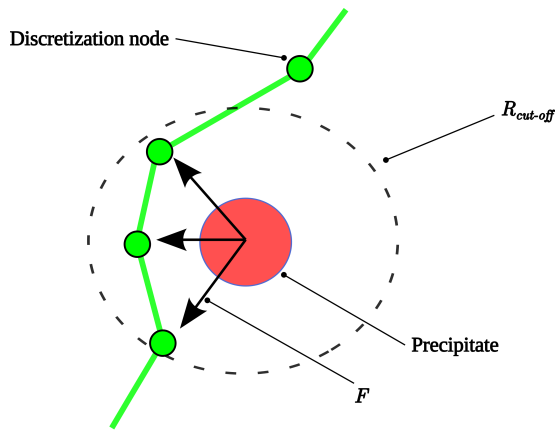}
\caption{(Color online). Schematic diagram of dislocation-precipitate interaction implementation in ParaDis. A force generated by the precipitate is applied to the discretization nodes which are inside the cut-off radius.}
\label{fig2}
\end{centering}
\end{figure}

The force from the precipitate is applied to the dislocation discretization nodes inside a cut-off radius 
$R_\text{cut-off}$, see Fig.~\ref{fig2}. In ParaDis, the discretization nodes can move along the dislocation line which introduces some numerical difficulties as new nodes are constantly generated to keep the line segments at certain length. This tangential node movement has no physical meaning, so we only use the precipitate force component which is normal to the segment for which the nodes are connected. This removes the unnecessary discretization operations and still preserves the physics of dislocation-precipitate interaction. When a dislocation segment is in the neighborhood of a 
precipitate, we need to make sure that its maximum length is of the same order or smaller 
than the size of the precipitate, in order to reduce numerical inaccuracy.

A common way of modeling the precipitate-dislocation 
interactions is to either assume that the dislocation movement ceases within the volume of the 
precipitate (impenetrable obstacle), or that the precipitate applies a constant 
drag force on the dislocation~\cite{mohles2004critical,monnet2011orowan}. In our model, the 
precipitates generate a spatially continuous force acting on the dislocations,
improving the numerical stability of the problem, and also taking in a crude fashion
into account the distance dependence of the elastic stress field of the point defect.

A full 3D system, where the dislocation can approach the obstacle from any direction, would 
require a more realistic model. In continuum elasticity the correct strain field for the 
obstacles is obtained from the Eshelby solution for spherical inclusions \cite{eshelby1957determination}. 
The force field of the obstacle would then have an angular as well as a radial dependence, 
and also include both attractive and repulsive components. This leads to a more complex 
interaction between the precipitate and the dislocation as it would also depend on the 
orientation of the dislocation. In our multi-scale model system, we have a single 
dislocation which is driven towards a precipitate situated at the glide plane of the dislocation. 
In this case the simple $r$ dependence of the Gaussian potential is sufficient to capture the 
essential physics of the dislocation pinning by the obstacle.
 
The obstacle strength is tunable [via $A$ and $R$ in Eq. (\ref{eq2})] which 
enables us to study both strong and weak pinning. Precipitates are treated similarly to 
the existing node data structure - for instance, precipitates can be removed or created during
the simulation. This also leads to the possibility of destructible pinning centers,
something that may be relevant in general e.g. for studies  of plastic instabilities in 
irradiated metals. 

\subsection{Multiscale framework}
When a dislocation driven by an applied stress encounters an immobile obstacle, 
it will become pinned. When the applied stress is increased to a critical value $\sigma_\text{c}$, 
the dislocation unpins (is able to move past the obstacle), and continues its 
movement. The nature  of the unpinning depends from the strength of the obstacle. 
When the obstacle is strong, unpinning happens via the Orowan-mechanism: with 
increasing stress, the dislocation bows around the obstacle and leaves an Orowan 
loop around it \cite{hull2011introduction}. The loop left behind increases the 
effective size of the obstacle. Thus, when multiple dislocations are driven through 
the obstacle, the loop formation process leads to strain hardening of the material:
$\sigma_\text{c}$ increases with each new loop. In this paper, we classify the 
obstacle as weak if the dislocation unpins without leaving a loop  behind, and consider
it to be strong if an Orowan loop is formed around the precipitate. Comparison of 
$\sigma_\text{c}$ in MD and DDD, respectively, allows us to fit the two methods 
together. In order to make  a realistic multiscale model, the input parameters for 
DDD were made as similar as possible to those of the MD simulations. From MD we can 
extract the shear modulus of the crystal, $G$, mobility of the dislocations, $M_\text{e}$ 
(where the subscript denotes edge), dislocation core energy, $E_{\text{core}}$, 
and the critical stress, $\sigma_\text{c}$, needed to overcome the precipitate. 

\section{Results}

In the following sections we will first present the results from the MD simulations and then the general properties of the DDD results. 
Finally  we will compare MD and DDD results in order to find suitable fitting parameters for the DDD precipitate potential. 

\subsection{Results from MD simulations}

\begin{figure}
\begin{centering}
\includegraphics[scale=0.35,clip=true,trim= 0 0 0 0]{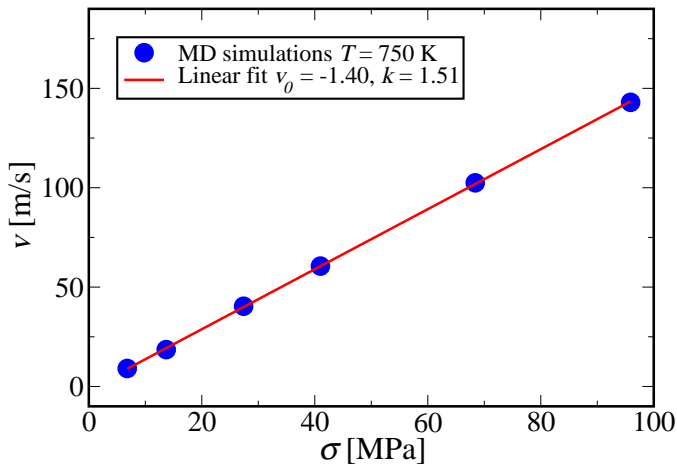}
\caption{(Color online). Stress-velocity data obtained from MD-simulations of iron at the temperature $ T=750 \, \text{K}$.  }
\label{fig3}
\end{centering}
\end{figure}

The parameters obtained in this paper and previously ~\cite{hen13} for the H13 
potential are listed in Tab.~\ref{table1}. The results 
from the 6 different stresses and the corresponding dislocation velocities are shown in 
Fig.~\ref{fig3}; a roughly linear velocity response to applied stresses is observed. 
From the linear fit to the data, we calculate the dislocation mobility needed in the DDD 
simulations.

\begin{figure}
\begin{centering}
\includegraphics[scale=0.32,clip=true,trim= 0 0 0 0]{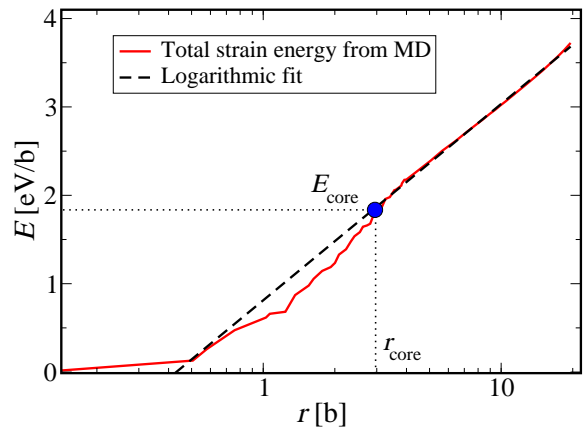}
\caption{(Color online).Total strain energy as function of the distance from dislocation core for a edge dislocation. Data is obtained from the MD-simulations. The dislocation core energy $E_{\text{core}} $, is the value of the total strain energy curve at the radius of the core. Outside of the the core, the total strain energy follows a logarithmic relation  $E\propto \ln{r}$ as predicted by elasticity theory. }
\label{fig5}
\end{centering}
\end{figure}

The data for Burgers vector $b$, the core energy $E_{\text{core}} $ and the core radius $r_{\text{core}}$ were obtained from 
Ref.~\cite{gra1}. Burgers vector is $0.2502$ nm, the core energy is $1.84\, \text{eV/b}$ and the core radius is $2.9\, \text{b}$. Elasticity theory predicts a logarithmic relation, $E\propto\ln{r}$, for total strain energy as function of the distance from the dislocation core \cite{hull2011introduction}. Using this fact, the core energy can be obtained from the total strain energy curve at the point where the strain energy starts to vary logarithmically (Fig.~\ref{fig5}). 

We calculated the shear modulus for the used potential at $750$ K to be 
about $75\,\text{GPa}$ (in $[111]$ direction) and the Poisson ratio at the same 
temperature to be $0.379$, by using the procedures described in section II.

To investigate how the unpinning stress depends on the spacing between obstacles, we 
investigated $7$ different lengths for the same $2\, \text{nm}$ fixed obstacle.
In the case of the constant strain rate $\dot{\gamma}=5\times10^{7}\,
\text{s}^{-1}$, we observed that the system responded with an extra 
flow stress before the dislocation started to move. To obtain the 
true strength of the obstacles we need to subtract this extra stress, 
to get only the contribution of the obstacle, not both the obstacle and 
the strain rate induced stress. The flow stress at the investigated 
temperature was observed to be $75\,\text{MPa}$, which has been subtracted 
from the obtained unpinning stresses. The unpinning stresses for the different lengths can be seen in Fig.~\ref{fig13} and in Fig.~\ref{fig14}, where the results from MD are compared with those of elasticity theory and the obtained results from DDD. An increasing spacing between obstacles
will decrease the needed unpinning stress, which is consistent with elasticity theory.

\begin{figure}
\begin{centering}
\includegraphics[scale=0.32,clip=true,trim= 0 0 0 0]{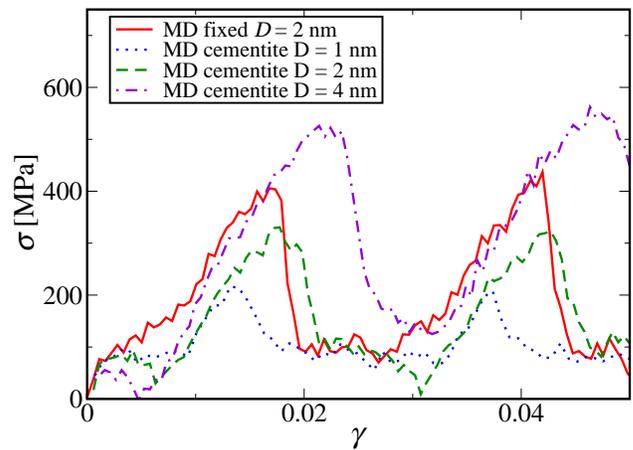}
\caption{(Color online). Stress-strain curves obtained from MD simulations of a dislocation interacting with a row of cementite precipitates of different sizes. Each stress drop corresponds to an unpinning event. The diameter of the precipitates are $D=1\,\text{nm}$, $2\,\text{nm}$ and   $4\,\text{nm}$ and the distance between obstacles is $L=21\, \text{nm} - D$.   }
\label{fig4}
\end{centering}
\end{figure}

The results of one of the cementite obstacles can be seen in 
Fig.~\ref{fig4}, where the stress strain curves for different obstacle sizes is shown. 
The figure shows that a larger obstacle will require a higher stress to 
unpin. In Fig. 6 the interaction of the edge dislocation with the 
$1\, \text{nm}$ cementite obstacle is shown. From the figure we see 
that the dislocation is pinned, but before the screw arms are created 
it unpins and can move past the obstacle. In Fig.~\ref{fig7} we see the interaction 
of the dislocation with a $4\, \text{nm}$ obstacle. Here the screw arms 
are created and extended, until the screw arms are attracted to each 
other and annihilate, letting the dislocation move past the obstacle.

\begin{table}
\caption{DDD simulation parameters obtained from MD simulations}
\label{table1}
\begin{center}
\begin{tabular}{ll}
Parameter	& Value   \\ \hline
$b$	& $0.2502\,\text{nm} $     \\
$r_{\text{core}}$	& $2.9 \,\text{b} $     \\
$E_{\text{core}}$& $1.84\, \dfrac{\text{eV}}{\text{b}}$ 						\\
$G$	& $75\, \text{GPa}$     \\
$\nu$	& $0.379$     \\
$M_{\text{edge}}$		&$6036.0\, (\text{Pa}\,\text{s})^{-1}$				\\
\end{tabular}
\end{center}
\end{table}

\begin{figure}
\begin{centering}
\includegraphics[scale=0.18,clip=true,trim= 4.5cm 0 0 0]{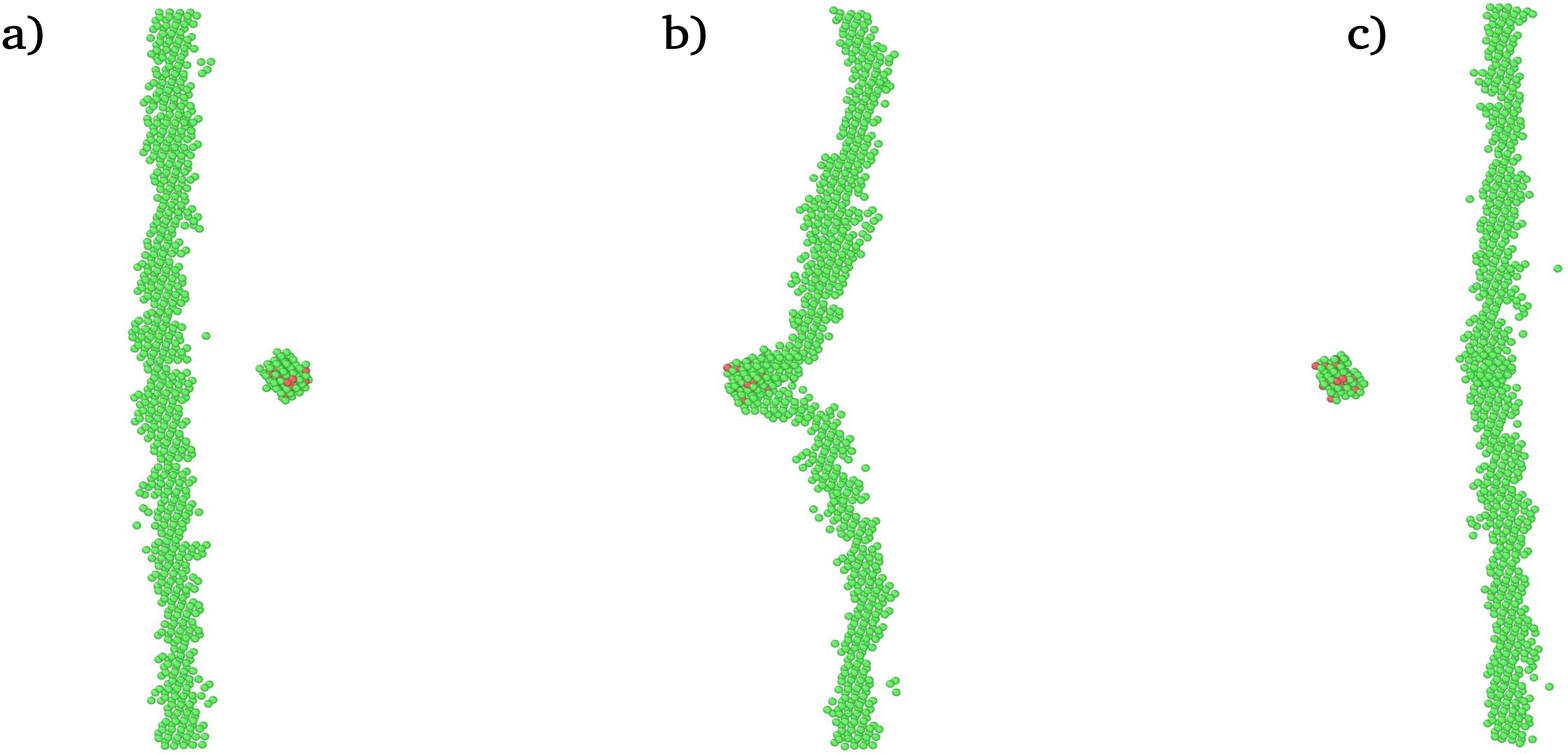}
\caption[]{(Color online). Dislocation interacting with a weak precipitate in the MD simulation. 
After some initial curving, the dislocation overcomes the precipitate potential and continues its movement. Simulation parameters are $L= 20.2\,\text{nm}$,$\dot{\gamma}=5\times10^{7}\,\text{s}^{-1}$ and $D= 1.0\,\text{nm}$ }
\end{centering}
\label{fig6}
\end{figure}


\begin{figure}
\centering
\subfigure[]{
\includegraphics[scale=0.095,clip=true,trim= 4.5cm 0 4.5cm 0]{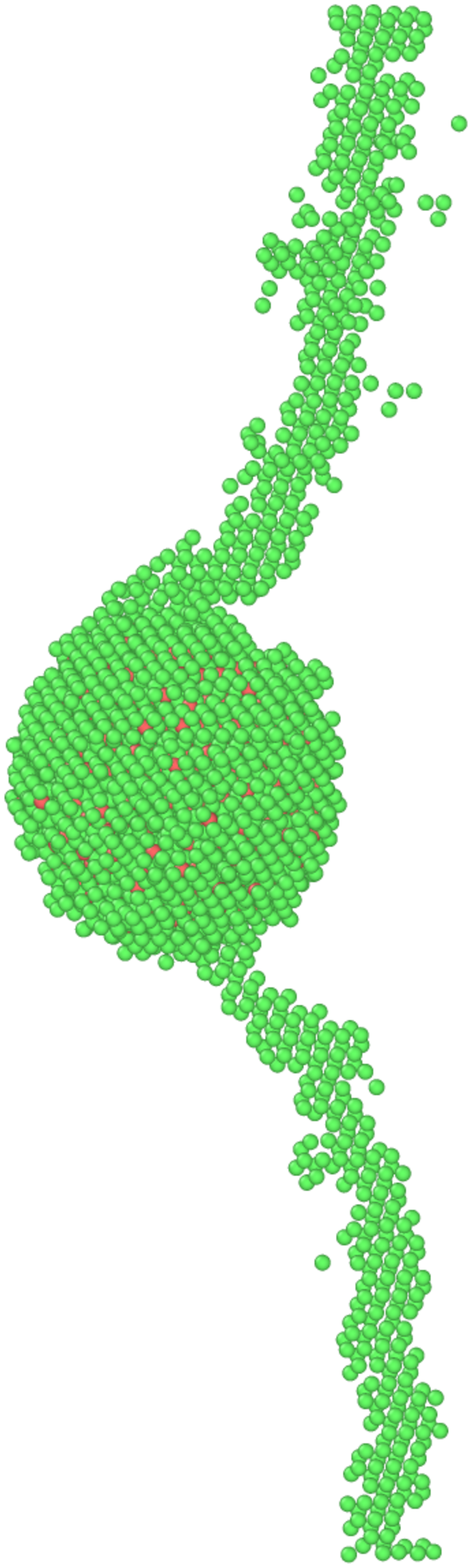}
\label{fig7a}
}
\subfigure[]{
\includegraphics[scale=0.095,clip=true,trim= 4.5cm 0  4.5cm 0]{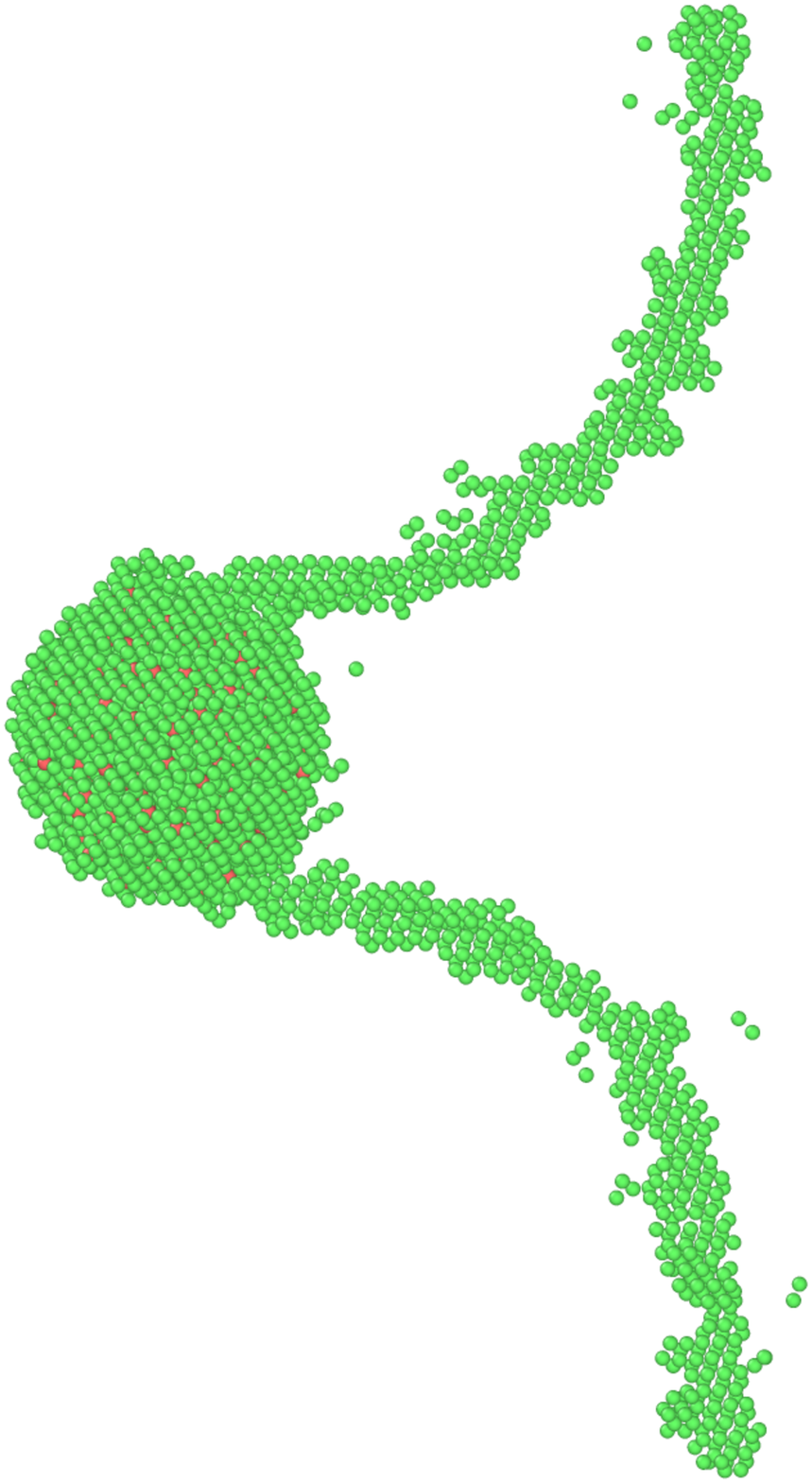}
\label{fig7b}
}
\\
\subfigure[]{
\includegraphics[scale=0.095,clip=true,trim= 4.5cm 0  4.5cm 0]{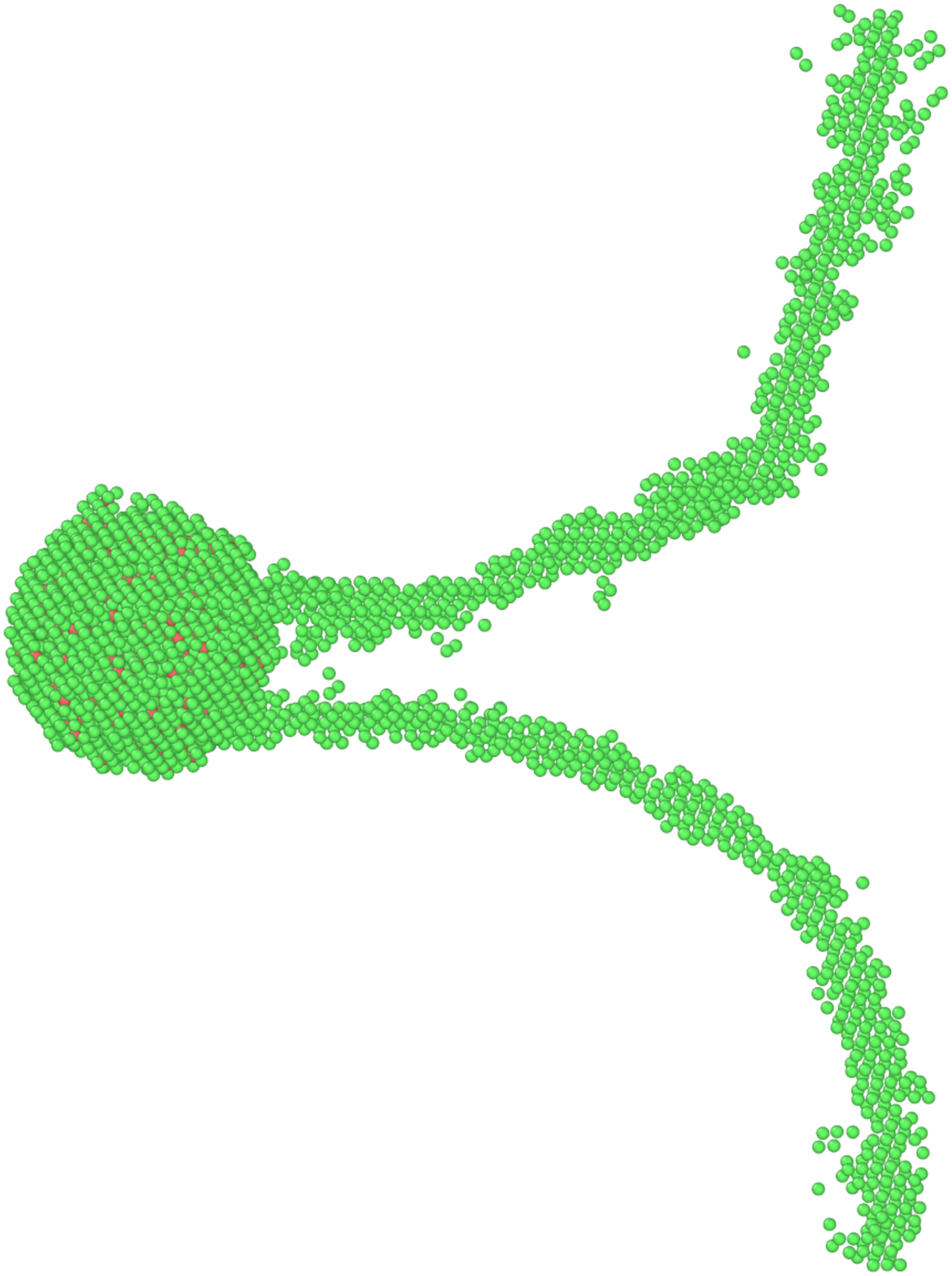}
\label{fig7c}
}
\subfigure[]{
\includegraphics[scale=0.095,clip=true,trim= 4.5cm 0  4.5cm 0]{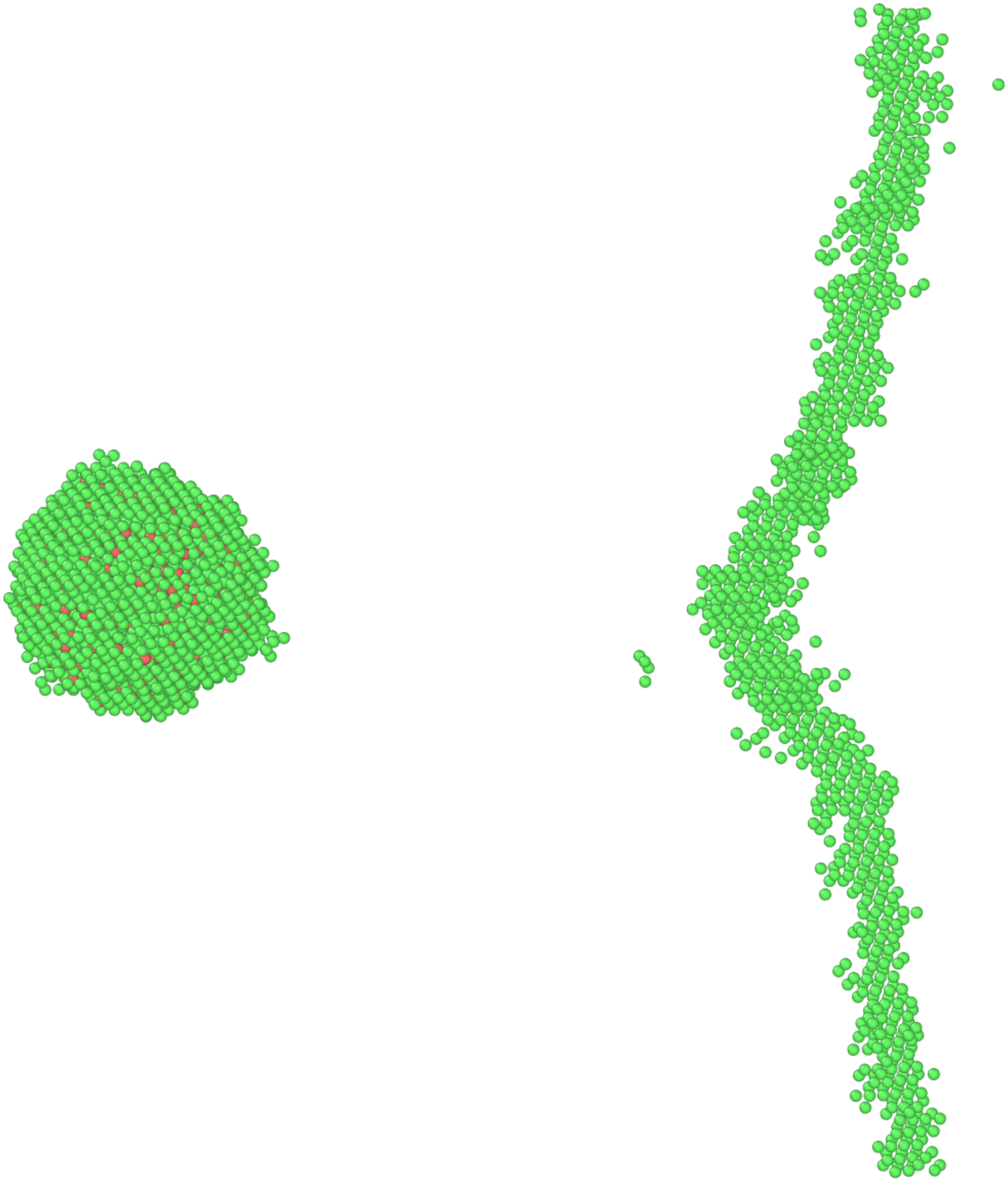}
\label{fig7d}
}

\caption[]{(Color online). Dislocation interacting with a strong precipitate in the MD simulation. Dislocation bypasses the precipitate by forming an Orowan loop around it. Simulation parameters are $L= 17.2\,\text{nm}$, $\dot{\gamma}=5\times10^{7}\,\text{s}^{-1}$ and $D=4.0\,\text{nm}$}
\label{fig7}
\end{figure}

\subsection{Results from DDD simulations}
In this work we use the default mobility $\textbf{BCC\_glide}$ of the ParaDis code,  in which the dislocation 
velocity is linearly proportional to the applied stress $\sigma$,
\begin{align}
\label{eq1}
v&= M_\text{e} b  \sigma\, , 
\end{align}
where $M_\text{e}$ is the edge mobility constant, and $b$ the Burgers vector of the 
dislocation. The choice of a linear mobility function is justified by the MD 
simulations at the temperature $T=750$ K; Fig.~\ref{fig3} shows that the velocity is 
linearly dependent on the applied stress. This temperature was chosen because it is typical for 
the operational conditions of steel structures in nuclear reactors, and 
because the shear modulus from the MD simulations for $T=750$ K is close to the corresponding 
experimental one~\cite{murty2008structural}. The ParaDis code is built for elastically isotropic materials such that e.g. their $G$ is 
the same in all directions. We use $G = 75\,\text{GPa}$ which is measured in the $[111]$ direction, 
perpendicular to the movement of the dislocation.
The edge dislocation mobility $M_\text{e}$ can be obtained from MD 
results via Eq. (\ref{eq1}), by making a linear fit to the stress-velocity data of 
Fig.~\ref{fig3}. We have done MD simulations only for edge dislocations; thus, the 
screw mobility remains a free parameter. Previous investigations have shown 
that screw mobility is about one third of the edge mobility at temperatures around
$T=750$ K, which we consider here \cite{queyreau2011edge,gilbert2011screw}. As a first 
order approximation, we assume that screw mobility is the same as the edge mobility, 
$M_\text{s}=M_\text{e}$. Tests of other choices are described below. The energy of the 
dislocation core, $E_\text{core}$, can 
be read from the total strain energy curve in Fig.~\ref{fig5}. The numerical values 
of all parameters are collected in Tab. \ref{table1}. In the calculation of the 
elastic energy of the dislocations, ParaDis uses a cut-off parameter $r_\text{core}$. 
This essentially tells the radius at which the core interactions replace linear elasticity theory. In MD 
simulations the core radius is $r_\text{core}=2.9 \, b$. 

In the case of small and strong pinning obstacles there were 
numerical problems in DDD simulations when using this value. These 
numerical problems arise when the distance between dislocation segments 
of the same Burgers vector equals or is smaller than the size of their 
core. For real dislocations the linear elasticity is no longer valid in 
this region and this is modeled in ParaDis by introducing a cut-off radius 
in the force calculation. This cut-off equals the  size of the  core radius 
$r_c$. Because of this cut-off, the force between dislocation segments is 
not strong enough to generate a stable configuration of layered Orowan 
loops. Dislocation segments start to partially merge, discretization nodes 
move in a random manner and time step shrinks orders of magnitude. We have 
used a smaller value $r_c=0.5\, b$ in order to overcome these numerical 
problems.

Here, we do not take into account possible dislocation climb or cross-slip; thus, we 
consider a dislocation which is constrained into its original glide plane throughout 
the simulation. In this way we can fit MD and DDD results together in the simplest 
scenario possible. We also assume that the precipitates are non-shearable and immobile. 
The dislocation may penetrate the obstacle but we do not remove it from the simulation after this. The default strain rate used is 
$\dot{\gamma}=10^{7}\,\text{s}^{-1} $.

\begin{figure}
\begin{centering}
\includegraphics[scale=0.18,clip=true,trim= 0 0 0 0]{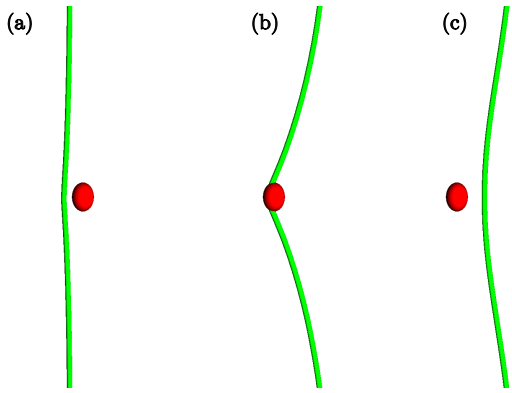}
\caption{(Color online). Dislocation interacting with a weak precipitate in the DDD simulation. 
After some initial curving, the dislocation overcomes the precipitate potential and continues its movement. Simulation parameters are $L= 42.5\,\text{nm}$, $R= 1.0\, \text{nm}$, $\dot{\gamma}=10^{7}\,\text{s}^{-1}$ and  $A=7.8\times 10^{-20}\, \text{Pa}\,\text{m}^3$.}
\label{fig8}
\end{centering}
\end{figure}

As stated previously, we categorize the precipitates as weak if they unpin without 
loop formation and strong if they leave a loop. A typical DDD simulation for weak 
pinning is presented in Fig.~\ref{fig8}. The edge dislocation is on the left side 
of the simulation box in the beginning, with the precipitate in the middle. Because 
of the periodic boundary conditions along the dislocation line direction, the 
dislocation  effectively sees an infinite row of precipitates. The distance between 
them, which is denoted by $L$, can be varied by changing the size of the simulation box. During simulations, 
the stress is increased in order to match the imposed strain rate, causing the 
dislocation to move right towards the precipitate. The dislocation then pins to the
precipitate. When the applied stress reaches a critical value 
$\sigma_\text{c}$, the dislocation is able to overcome the Gaussian potential and 
unpins from the precipitate.

\begin{figure}
\centering
\subfigure[]{
\includegraphics[scale=0.1,clip=true,trim= 0 0  0 0]{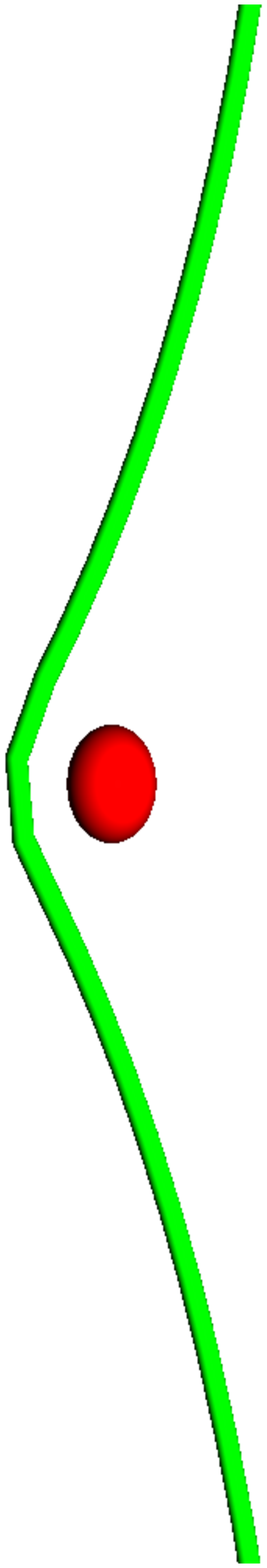}
\label{fig9a}
}
\subfigure[]{
\includegraphics[scale=0.1,clip=true,trim= 0  0  0 0]{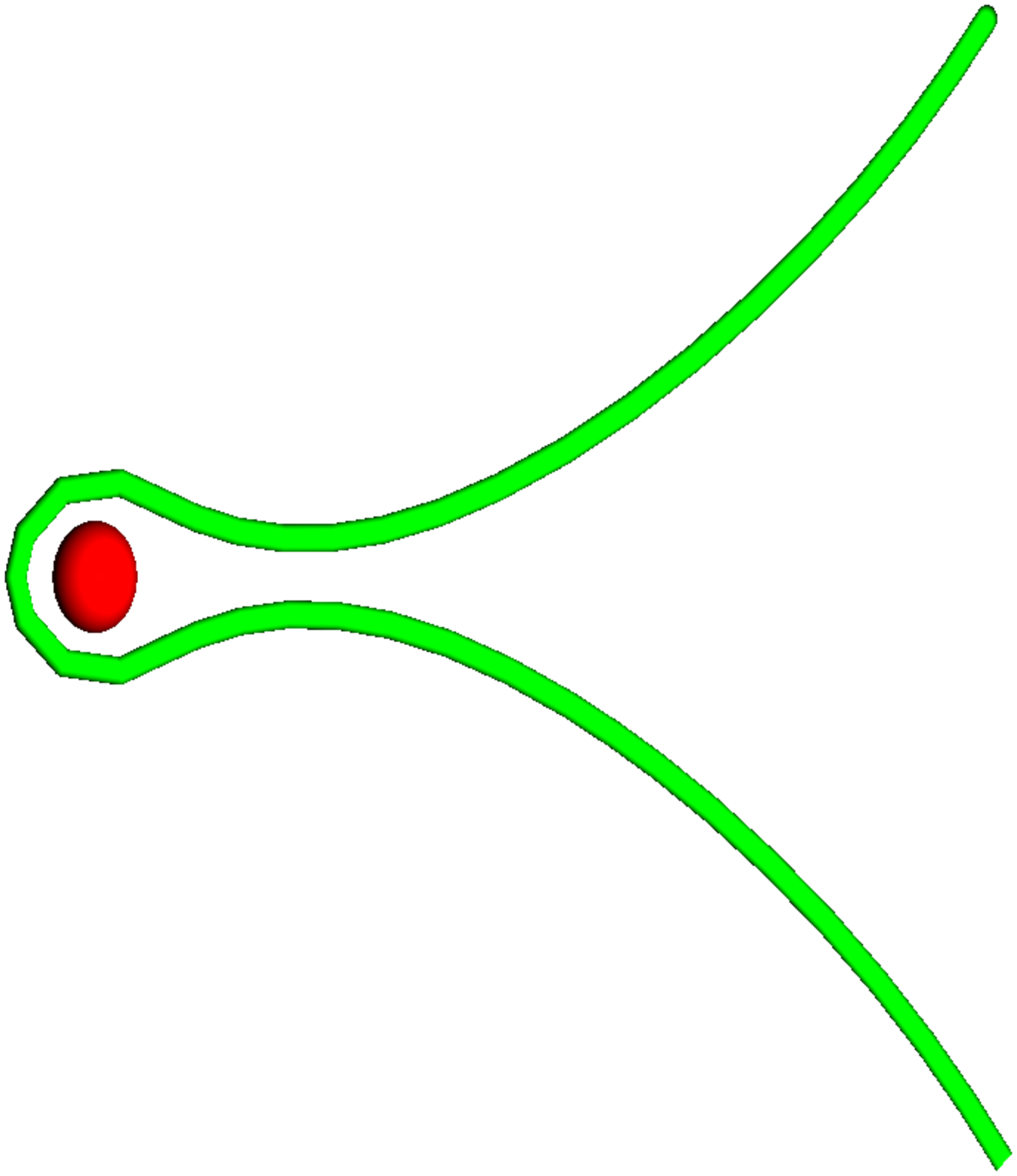}
\label{fig9b}
}
\\
\subfigure[]{
\includegraphics[scale=0.1,clip=true,trim= 0 0 0 0]{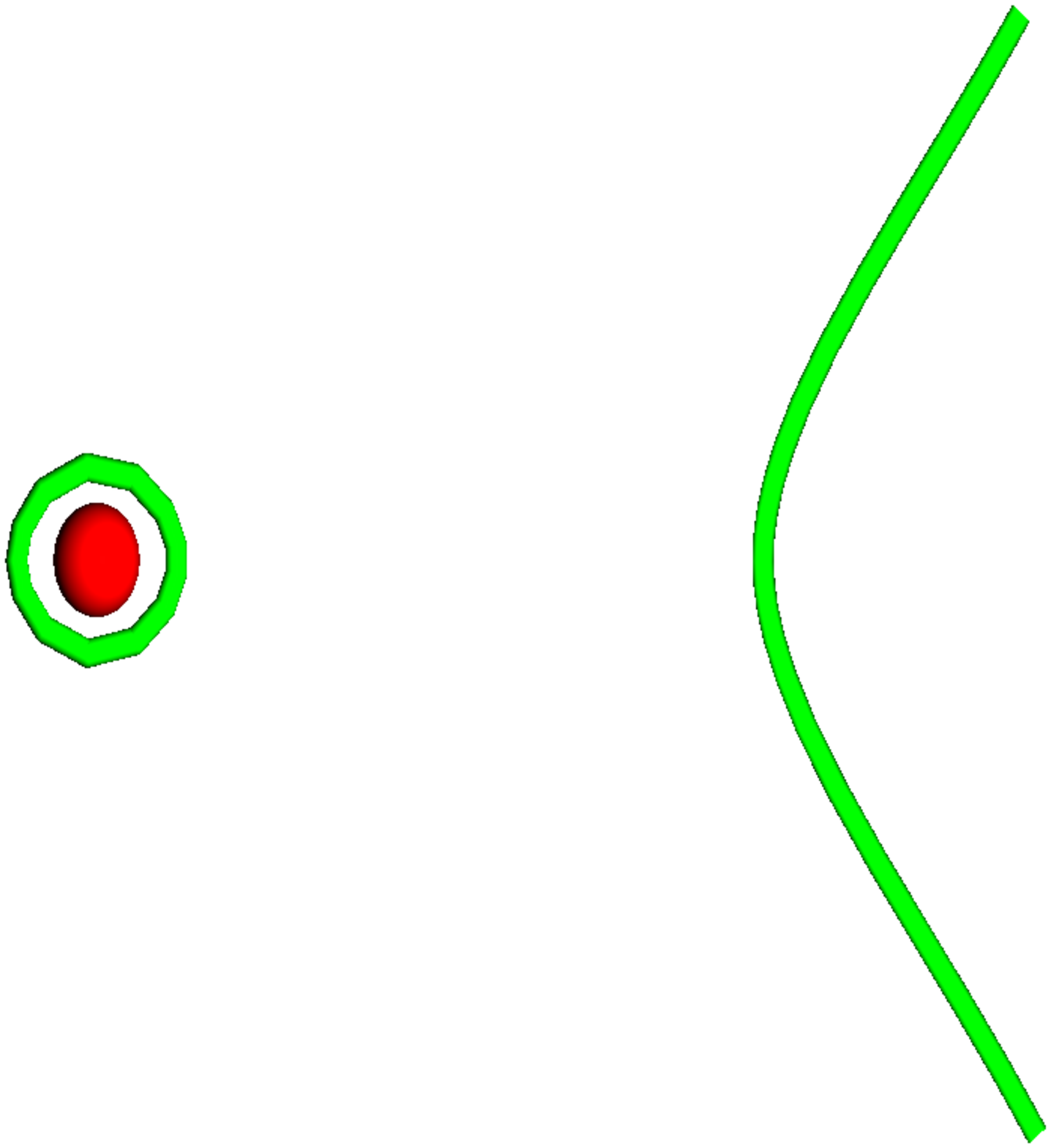}
\label{fig9c}
}
\subfigure[]{
\includegraphics[scale=0.1,clip=true,trim= 0 0 0 0]{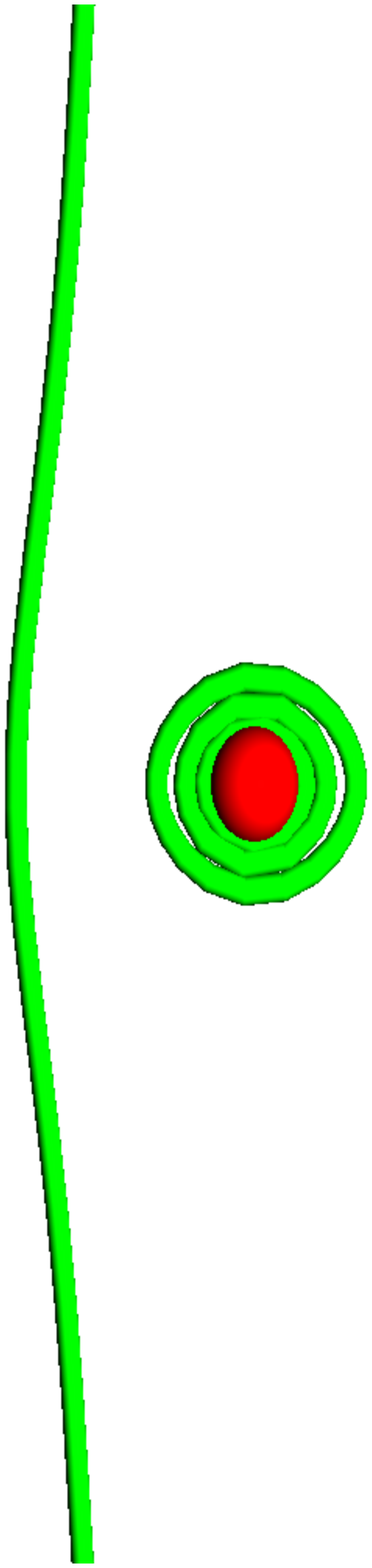}
\label{fig9d}
}

\caption[]{(Color online). Dislocation interacting with a strong precipitate in the DDD simulation. Dislocation bypasses the precipitate by forming an Orowan loop around it. Simulation parameters are $L= 42.5\,\text{nm}$, $\dot{\gamma}=10^{7}\,\text{s}^{-1}$, $R=1.0\,\text{nm}$ and $A=1.56\times10^{-18}\, \text{Pa}\,\text{m}^3$.}
\label{fig9}
\end{figure}

The case of strong pinning is presented in Fig.~\ref{fig9}. The edge dislocation is 
on the left side of the simulation box in the beginning. The stress is increased 
in order to match the imposed strain rate, which causes the dislocation to move 
right towards the precipitate. The dislocation then pins to the precipitate, Fig.~\ref{fig9a}. When stress reaches $\sigma_\text{c}$, the dislocation bows around 
the precipitate, leaving behind an Orowan loop [Figs.~\ref{fig9b} and \ref{fig9c}]. 
After many dislocations have been driven trough the simulation box, the precipitate 
has collected multiple loops around it [Fig.~\ref{fig9d}]. 

\begin{figure}
\begin{centering}
\includegraphics[scale=0.32,clip=true,trim= 0 0 0 0]{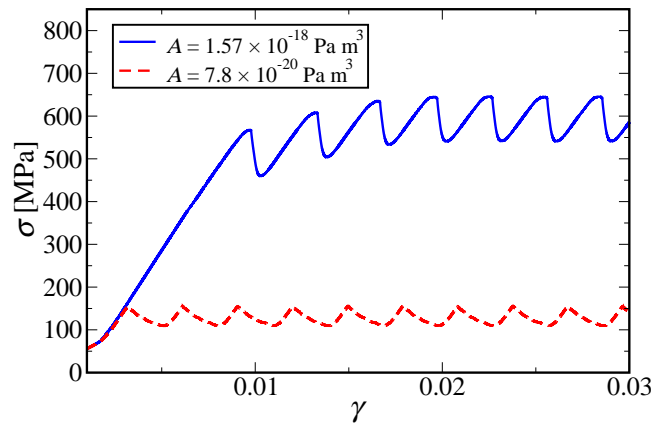}
\caption{(Color online). Stress-strain curves of the dislocation-precipitate interaction from DDD simulations. Continuous curve represents the strong precipitates and the dashed line represents the weak ones. Simulation parameters are  $L= 42.5\,\text{nm}$, $\dot{\gamma}=10^{7}\,\text{s}^{-1}$  and $ \,R=1.0\,\text{nm}$.  }
\label{fig10}
\end{centering}
\end{figure}

When the dislocation is pinned, the crystal strains essentially elastically until 
the dislocation bows out between the precipitates. This generates the distinctive 
serrated look of the stress-strain curves of Fig.~\ref{fig10}.
Each stress drop is related to an unpinning event in both cases. In the case of the strong obstacle, 
the precipitate gather loops around it until the stress between loop segments is 
large enough to collapse the inner loop. After this, the precipitate is surrounded by 
a constant number of loops, and consequently also $\sigma_\text{c}$ becomes a constant.  
This can be seen in the stress-strain curve of the strong precipitate in Fig.~\ref{fig10}. The stress drops are increasing in height until the third drop. This 
means that the precipitate has a maximum of three loops around it.

%
%
%

\begin{figure}
\begin{centering}
\includegraphics[scale=0.32,clip=true,trim= 0 0 0 0]{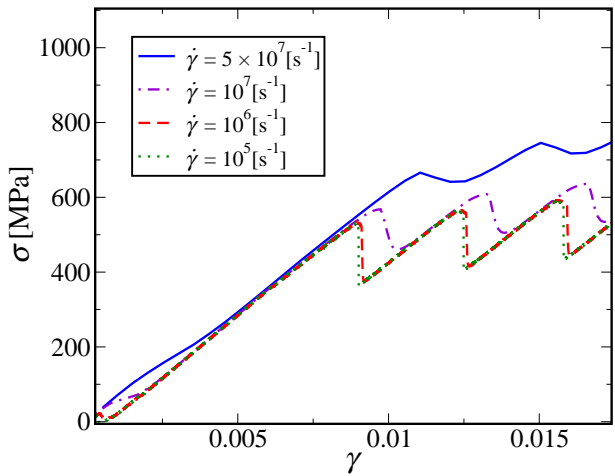}
\caption{(Color online). Stress-strain curves with different shear-rates for strong 
precipitates. The unpinning stress saturates when the strain rate is decreased. Simulation parameters are $L=42.5\,\text{nm}$, $A=1.56\times10^{-18}\, \text{Pa} \, \text{m}^3\,$and $R=1.0\,\text{nm}$. }
\label{fig11}
\end{centering}
\end{figure}

\begin{figure}
\begin{centering}
\includegraphics[scale=0.32,clip=true,trim= 0 0 0 0]{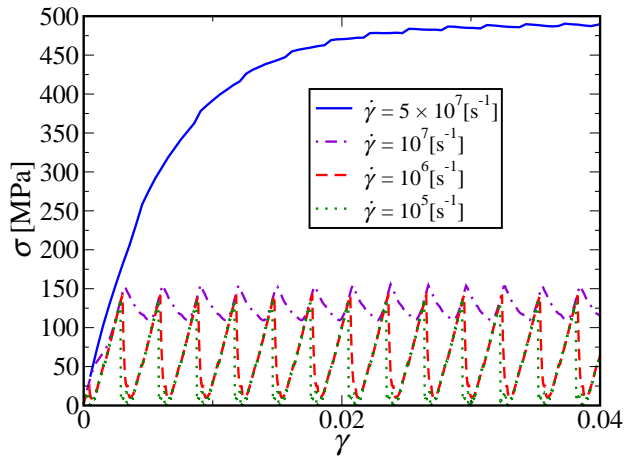}
\caption{(Color online). Stress-strain curves with different shear-rates for weak 
precipitates. The unpinning stress saturates when the strain rates is decreased. Distance between the precipitates is $L=42.5\,\text{nm}$. Simulation parameters are $A=7.8\times10^{-20}\, \text{Pa} \, \text{m}^3$, and $R=1.0\,\text{nm}$. }
\label{fig12}
\end{centering}
\end{figure}

In DDD simulations we can use strain rates that are orders of magnitude 
smaller than in MD. As Figs.~\ref{fig11} and ~\ref{fig12} indicate, the stress-strain 
curves look qualitatively similar with all imposed strain rates, but the magnitude of $\sigma_\text{c}$ gets smaller when the strain rate is decreased. This dependence of $\sigma_\text{c}$ on $\dot{\gamma}$ is smaller for lower strain rates, $\dot{\gamma}=10^{6}\,\text{s}^{-1}$ 
and $\dot{\gamma}=10^{5}\,\text{s}^{-1}$. This behavior is possibly due to the low dislocation density, as there is only a single dislocation in the simulation space. When the dislocation density is low and the imposed strain rate high, the crystal strains mostly elastically as the movement of the dislocation does not produce plastic strain fast enough to satisfy the high strain rate even when it's not pinned by a precipitate. With lower strain rates, the dislocation has time to react to the imposed stress, and to produce the imposed strain rate plastically after it unpins from the precipitate.

As stated previously, we do not have MD results for the mobility of 
screw dislocations. Thus, we have assumed that it has the same value as the edge mobility. 
We check the validity of this assumption by decreasing the magnitude of the 
screw mobility, while keeping the edge mobility constant. When the screw mobility 
is of the order $M_\text{s}\approx0.01 \cdot M_\text{e}$, there is a qualitatively different bow-out 
during unpinning. The screw segments form a long dipole after the precipitate before 
they annihilate, and an Orowan loop is formed. This effect, however, does not change 
$\sigma_\text{c}$ significantly. This result is supported by previous DDD studies 
by Monnet\textit{ et al}., whose simulations show that the effect of the different mobilities
on $\sigma_\text{c}$ should be small in the range of the $D/L$ ratio studied here \cite{monnet2011orowan}.

\subsection{Comparison of MD and DDD}

In order to find good fitting parameters for the Gaussian potential, we compare 
the $\sigma_\text{c}$ from MD simulations of fixed obstacles to the ones obtained from DDD simulations when using the same strain rate $\dot{\gamma}=5\times10^{7}\,\text{s}^{-1}$. 
In this comparison the critical stress is defined as the first stress drop of the respective
stress-strain curves. Both MD and DDD are then compared to the Bacon-Kocks-Scattergood 
(BKS) equation \cite{bacon1973effect,monnet2011orowan}
\begin{align}
\label{eq3}
\sigma_c &=C\frac{G b}{L-D}\left[\ln\left(\dfrac{\bar{D}}{b}\right)+0.7\right]  \,, \nonumber\\
\end{align}
where for edge dislocations $C=\frac{1}{2\pi}$, $L$ is the distance between obstacles,  
$D$ is the diameter of the obstacles, and $\bar{D}=\dfrac{DL}{D+L}$ is the harmonic 
average of $L$ and $D$. This formula is obtained by considering only the dislocation self-interaction in the case where the dislocation is curved around an infinitely hard exactly spherical obstacle. This approximation differs from the MD and DDD simulations where there is a continuous stress-field around the obstacles. The BKS equation gives larger values of $\sigma_\text{c}$ than MD which can be explained by the precipitates in 
MD being penetrable - i.e. the dislocations can bypass them without leaving loops behind 
at the temperature of $750$ K. This penetration can be due to a climb or a cross-slip 
process, not considered in the DDD simulations.   

\begin{figure}
\begin{centering}
\includegraphics[scale=0.32,clip=true,trim= 0 0 0 0]{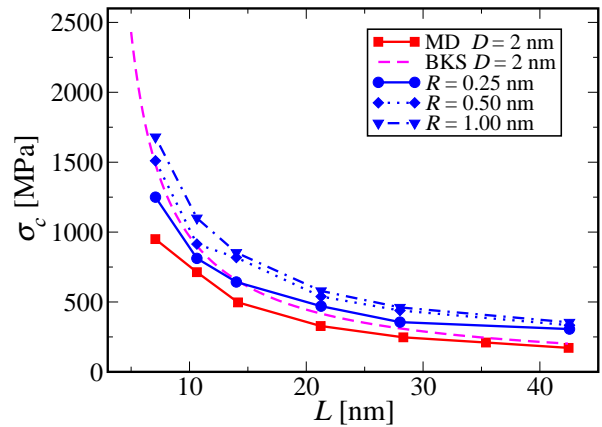}
\caption{(Color online). Critical stress $\sigma_c$  as a function of the distance between precipitates $L$ for different precipitate sizes $R$. Continuous curve with square symbols denotes results from MD simulations and the dashed curve from the BKS equation. The rest are results from DDD simulations. DDD Simulation parameters are $A=1.56\times10^{-19}\,\text{Pa}\,\text{m}^3$ and  $\dot{\gamma}=5\times10^{7}\,\text{s}^{-1}$. }
\label{fig13}
\end{centering}
\end{figure}

The critical stress as a function of the precipitate size is presented in Fig.~\ref{fig13}. The critical stress increases with the size of the precipitates.  

\begin{figure}
\begin{centering}
\includegraphics[scale=0.32,clip=true,trim= 0 0 0 0]{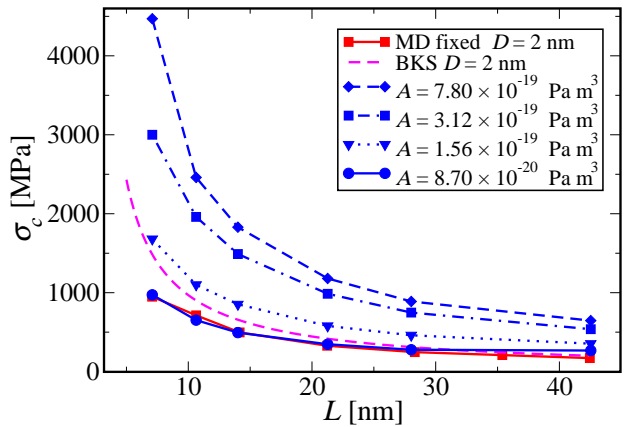}
\caption{(Color online). Critical stress $\sigma_c$ as a function of the distance between precipitates $L$ for different precipitate strengths $A$. Continuous curve with square symbols denotes results from MD simulations and the dashed curve from the BKS equation. The rest are results from DDD simulations. DDD simulation parameters are $R=1.0\,\text{nm}$ and $\dot{\gamma}=5\times10^{7}\,\text{s}^{-1}$.}
\label{fig14}
\end{centering}
\end{figure}

The critical stress  as a function of the precipitate strength is presented in Fig.~\ref{fig14}. Strong precipitates are represented by the dotted and dashed lines, and the weak precipitates by the continuous line.

A good fit between MD and DDD results is obtained with precipitate strength parameter value $A=8.7\times10^{-20}\,\text{Pa}\,\text{m}^3$, which corresponds to a weak precipitate. This means that there is no Orowan-loop formation. The MD stress-strain curves in Fig.~\ref{fig4} supports this result as the height of the stress drops is not increased significantly when multiple dislocations are driven trough the system. We can also see from Figs.~\ref{fig13} and~\ref{fig14} that when the distance between the pinning points is large compared to their diameter $D/L\ll 1 $, the details of the dislocation-precipitate interaction do not change the unpinning stress .

\section{Discussion and Conclusions}
The critical stress obtained from MD simulations is smaller 
than that predicted by the BKS equation. This may be due to dislocation climb and/or 
cross-slip, as BKS does not take these into account. BKS also assumes that the 
precipitates are impenetrable, and exactly spherical obstacles with clear edges. 
Our Gaussian potential on the other hand generates a continuous force field and 
thus the edge of the obstacle is not well-defined. When a dislocation moves towards 
its center, the effective radius of the obstacle becomes smaller and this decreases 
the critical stress. In MD simulations the deformation of the precipitate can also be an 
important factor, which is not addressed in the current DDD implementation.

The DDD model can be fitted to match both the MD and BKS by varying the pinning 
strength parameter of the potential. With small $A$, the dislocation penetrates the 
precipitate and no Orowan loops are formed. With a large enough $A$, the 
dislocation bypasses the precipitate by leaving Orowan loops around the
obstacle, leading to Orowan hardening after the precipitate has gathered multiple 
loops around it. This kind of behavior is not likely to be captured with models 
which use impenetrable obstacles or ones with a constant drag force.
 
There are, however, some restrictions. In DDD simulations one must use a cubic simulation box 
as the code uses spatial symmetries in far field calculations by assuming a cubic 
simulation space. Because of this restriction, the dislocation densities are not 
the same in the two simulations, leading to different accumulated strains. 
This difference does not affect significantly the magnitude of  $\sigma_\text{c}$ at low strain rates. We were able to obtain good fit between MD and DDD results. The results indicate that the dislocation does not leave an Orowan loop around the precipitate at the temperature of 750 \text{K}. 

With our model it is easy to tune the strength of the precipitates. This 
offers possibilities to investigate dislocation systems with frozen disorder 
where the magnitude of the disorder is a controllable parameter.
For example the effect of pinning points to dislocation avalanches have been studied in 2D \cite{ovaska2014collective} where it was found that the presence of defects changed the statistics of avalanches compared to those in a pure dislocation system. It would be interesting to study if this would be the case also in a 3D dislocation system.  

Another area of application would be the strain hardening of irradiated metals.
This could be studied in a system where the size and strength of the precipitates would follow a realistic distribution obtained from experimental material microstructure data.
The effects of cross-slipping and dislocation climb on $\sigma_c$ are also a straightforward venue for future research. A more realistic model for the stress field of the precipitate is possible with the Eshelby-solution for the spherical inclusion \cite{eshelby1957determination}. This would lead to a physically more accurate model for the precipitate-dislocation interaction, which could then be compared to existing MD results \cite{gra2,gra3}.

\begin{acknowledgments}
This work has been supported by Academy of Finland via the SIRDAME project 
(project nos. 259886 and 260053), through the Centres of Excellence program (2012-2017, 
project no. 251748), and through an Academy Research Fellowship (LL, project no. 268302).
This work has been carried out within the framework of the EUROfusion Consortium and has received funding from the Euratom research and training programme 2014-2018 under grant agreement No 633053. The views and opinions expressed herein do not necessarily reflect those of the European Commission.
 We acknowledge the computational resources provided by the Aalto University 
School of Science ``Science-IT'' project, as well as those provided by CSC (Finland).
\end{acknowledgments}

\bibliography{manuscript}

\begin{thebibliography}{29}%
\makeatletter
\providecommand \@ifxundefined [1]{%
 \@ifx{#1\undefined}
}%
\providecommand \@ifnum [1]{%
 \ifnum #1\expandafter \@firstoftwo
 \else \expandafter \@secondoftwo
 \fi
}%
\providecommand \@ifx [1]{%
 \ifx #1\expandafter \@firstoftwo
 \else \expandafter \@secondoftwo
 \fi
}%
\providecommand \natexlab [1]{#1}%
\providecommand \enquote  [1]{``#1''}%
\providecommand \bibnamefont  [1]{#1}%
\providecommand \bibfnamefont [1]{#1}%
\providecommand \citenamefont [1]{#1}%
\providecommand \href@noop [0]{\@secondoftwo}%
\providecommand \href [0]{\begingroup \@sanitize@url \@href}%
\providecommand \@href[1]{\@@startlink{#1}\@@href}%
\providecommand \@@href[1]{\endgroup#1\@@endlink}%
\providecommand \@sanitize@url [0]{\catcode `\\12\catcode `\$12\catcode
  `\&12\catcode `\#12\catcode `\^12\catcode `\_12\catcode `\%12\relax}%
\providecommand \@@startlink[1]{}%
\providecommand \@@endlink[0]{}%
\providecommand \url  [0]{\begingroup\@sanitize@url \@url }%
\providecommand \@url [1]{\endgroup\@href {#1}{\urlprefix }}%
\providecommand \urlprefix  [0]{URL }%
\providecommand \Eprint [0]{\href }%
\providecommand \doibase [0]{http://dx.doi.org/}%
\providecommand \selectlanguage [0]{\@gobble}%
\providecommand \bibinfo  [0]{\@secondoftwo}%
\providecommand \bibfield  [0]{\@secondoftwo}%
\providecommand \translation [1]{[#1]}%
\providecommand \BibitemOpen [0]{}%
\providecommand \bibitemStop [0]{}%
\providecommand \bibitemNoStop [0]{.\EOS\space}%
\providecommand \EOS [0]{\spacefactor3000\relax}%
\providecommand \BibitemShut  [1]{\csname bibitem#1\endcsname}%
\let\auto@bib@innerbib\@empty
\bibitem [{\citenamefont {Terentyev}\ \emph {et~al.}(2008)\citenamefont
  {Terentyev}, \citenamefont {Bacon},\ and\ \citenamefont
  {Osetsky}}]{terentyev2008interaction}%
  \BibitemOpen
  \bibfield  {author} {\bibinfo {author} {\bibfnamefont {D.}~\bibnamefont
  {Terentyev}}, \bibinfo {author} {\bibfnamefont {D.~J.}\ \bibnamefont
  {Bacon}}, \ and\ \bibinfo {author} {\bibfnamefont {Y.~N.}\ \bibnamefont
  {Osetsky}},\ }\href@noop {} {\bibfield  {journal} {\bibinfo  {journal} {J.
  Phys. Cond. Matter}\ }\textbf {\bibinfo {volume} {20}},\ \bibinfo {pages}
  {445007} (\bibinfo {year} {2008})}\BibitemShut {NoStop}%
\bibitem [{\citenamefont {De~Koning}\ \emph {et~al.}(2003)\citenamefont
  {De~Koning}, \citenamefont {Kurtz}, \citenamefont {Bulatov}, \citenamefont
  {Henager}, \citenamefont {Hoagland}, \citenamefont {Cai},\ and\ \citenamefont
  {Nomura}}]{Koning2003modeling}%
  \BibitemOpen
  \bibfield  {author} {\bibinfo {author} {\bibfnamefont {M.}~\bibnamefont
  {De~Koning}}, \bibinfo {author} {\bibfnamefont {R.~J.}\ \bibnamefont
  {Kurtz}}, \bibinfo {author} {\bibfnamefont {V.~V.}\ \bibnamefont {Bulatov}},
  \bibinfo {author} {\bibfnamefont {C.~H.}\ \bibnamefont {Henager}}, \bibinfo
  {author} {\bibfnamefont {R.~G.}\ \bibnamefont {Hoagland}}, \bibinfo {author}
  {\bibfnamefont {W.}~\bibnamefont {Cai}}, \ and\ \bibinfo {author}
  {\bibfnamefont {M.}~\bibnamefont {Nomura}},\ }\href@noop {} {\bibfield
  {journal} {\bibinfo  {journal} {J. Nucl. Mater.}\ }\textbf {\bibinfo {volume}
  {323}},\ \bibinfo {pages} {281} (\bibinfo {year} {2003})}\BibitemShut
  {NoStop}%
\bibitem [{\citenamefont {de~la Rubia}\ \emph {et~al.}(2000)\citenamefont
  {de~la Rubia}, \citenamefont {Zbib}, \citenamefont {Khraishi}, \citenamefont
  {Wirth}, \citenamefont {Victoria},\ and\ \citenamefont
  {Caturla}}]{Rubia2000multiscale}%
  \BibitemOpen
  \bibfield  {author} {\bibinfo {author} {\bibfnamefont {T.~D.}\ \bibnamefont
  {de~la Rubia}}, \bibinfo {author} {\bibfnamefont {H.~M.}\ \bibnamefont
  {Zbib}}, \bibinfo {author} {\bibfnamefont {T.~A.}\ \bibnamefont {Khraishi}},
  \bibinfo {author} {\bibfnamefont {B.~D.}\ \bibnamefont {Wirth}}, \bibinfo
  {author} {\bibfnamefont {M.}~\bibnamefont {Victoria}}, \ and\ \bibinfo
  {author} {\bibfnamefont {M.~J.}\ \bibnamefont {Caturla}},\ }\href@noop {}
  {\bibfield  {journal} {\bibinfo  {journal} {Nature}\ }\textbf {\bibinfo
  {volume} {406}},\ \bibinfo {pages} {871} (\bibinfo {year}
  {2000})}\BibitemShut {NoStop}%
\bibitem [{\citenamefont {Wirth}\ \emph {et~al.}(2004)\citenamefont {Wirth},
  \citenamefont {Odette}, \citenamefont {Marian}, \citenamefont {Ventelon},
  \citenamefont {Young-Vandersall},\ and\ \citenamefont
  {Zepeda-Ruiz}}]{wirth2004multiscale}%
  \BibitemOpen
  \bibfield  {author} {\bibinfo {author} {\bibfnamefont {B.~D.}\ \bibnamefont
  {Wirth}}, \bibinfo {author} {\bibfnamefont {G.~R.}\ \bibnamefont {Odette}},
  \bibinfo {author} {\bibfnamefont {J.}~\bibnamefont {Marian}}, \bibinfo
  {author} {\bibfnamefont {L.}~\bibnamefont {Ventelon}}, \bibinfo {author}
  {\bibfnamefont {J.~A.}\ \bibnamefont {Young-Vandersall}}, \ and\ \bibinfo
  {author} {\bibfnamefont {L.~A.}\ \bibnamefont {Zepeda-Ruiz}},\ }\href@noop {}
  {\bibfield  {journal} {\bibinfo  {journal} {J. Nucl. Mater.}\ }\textbf
  {\bibinfo {volume} {329}},\ \bibinfo {pages} {103} (\bibinfo {year}
  {2004})}\BibitemShut {NoStop}%
\bibitem [{\citenamefont {Dewald}\ and\ \citenamefont
  {Curtin}(2007)}]{dewald2007multiscale}%
  \BibitemOpen
  \bibfield  {author} {\bibinfo {author} {\bibfnamefont {M.~P.}\ \bibnamefont
  {Dewald}}\ and\ \bibinfo {author} {\bibfnamefont {W.~A.}\ \bibnamefont
  {Curtin}},\ }\href@noop {} {\bibfield  {journal} {\bibinfo  {journal}
  {Modell. Simul. Mater. Sci. Eng.}\ }\textbf {\bibinfo {volume} {15}},\
  \bibinfo {pages} {S193} (\bibinfo {year} {2007})}\BibitemShut {NoStop}%
\bibitem [{\citenamefont {Kumar}\ \emph {et~al.}(2012)\citenamefont {Kumar},
  \citenamefont {Durgaprasad}, \citenamefont {Dutta},\ and\ \citenamefont
  {Dey}}]{kumar2012multiscale}%
  \BibitemOpen
  \bibfield  {author} {\bibinfo {author} {\bibfnamefont {N.~N.}\ \bibnamefont
  {Kumar}}, \bibinfo {author} {\bibfnamefont {P.~V.}\ \bibnamefont
  {Durgaprasad}}, \bibinfo {author} {\bibfnamefont {B.~K.}\ \bibnamefont
  {Dutta}}, \ and\ \bibinfo {author} {\bibfnamefont {G.~K.}\ \bibnamefont
  {Dey}},\ }\href@noop {} {\bibfield  {journal} {\bibinfo  {journal} {Comput.
  Mater. Sci.}\ }\textbf {\bibinfo {volume} {53}},\ \bibinfo {pages} {258}
  (\bibinfo {year} {2012})}\BibitemShut {NoStop}%
\bibitem [{\citenamefont {Bulatov}\ and\ \citenamefont
  {Cai}(2006)}]{bulatov2006computer}%
  \BibitemOpen
  \bibfield  {author} {\bibinfo {author} {\bibfnamefont {V.}~\bibnamefont
  {Bulatov}}\ and\ \bibinfo {author} {\bibfnamefont {W.}~\bibnamefont {Cai}},\
  }\href@noop {} {\emph {\bibinfo {title} {Computer simulations of
  dislocations}}},\ Vol.~\bibinfo {volume} {3}\ (\bibinfo  {publisher} {Oxford
  University Press},\ \bibinfo {year} {2006})\BibitemShut {NoStop}%
\bibitem [{\citenamefont {Groh}\ and\ \citenamefont
  {Zbib}(2009)}]{groh2009multiscale}%
  \BibitemOpen
  \bibfield  {author} {\bibinfo {author} {\bibfnamefont {S.}~\bibnamefont
  {Groh}}\ and\ \bibinfo {author} {\bibfnamefont {H.~M.}\ \bibnamefont
  {Zbib}},\ }\href@noop {} {\bibfield  {journal} {\bibinfo  {journal} {J. Eng.
  Mater. Technol.}\ }\textbf {\bibinfo {volume} {131}},\ \bibinfo {pages}
  {041209} (\bibinfo {year} {2009})}\BibitemShut {NoStop}%
\bibitem [{\citenamefont {Odette}\ \emph {et~al.}(2008)\citenamefont {Odette},
  \citenamefont {Alinger},\ and\ \citenamefont {Wirth}}]{odette2008recent}%
  \BibitemOpen
  \bibfield  {author} {\bibinfo {author} {\bibfnamefont {G.~R.}\ \bibnamefont
  {Odette}}, \bibinfo {author} {\bibfnamefont {M.~J.}\ \bibnamefont {Alinger}},
  \ and\ \bibinfo {author} {\bibfnamefont {B.~D.}\ \bibnamefont {Wirth}},\
  }\href@noop {} {\bibfield  {journal} {\bibinfo  {journal} {Annu. Rev. Mater.
  Res.}\ }\textbf {\bibinfo {volume} {38}},\ \bibinfo {pages} {471} (\bibinfo
  {year} {2008})}\BibitemShut {NoStop}%
\bibitem [{\citenamefont {Hirata}\ \emph {et~al.}(2011)\citenamefont {Hirata},
  \citenamefont {Fujita}, \citenamefont {Wen}, \citenamefont {Schneibel},
  \citenamefont {Liu},\ and\ \citenamefont {Chen}}]{hirata2011}%
  \BibitemOpen
  \bibfield  {author} {\bibinfo {author} {\bibfnamefont {A.}~\bibnamefont
  {Hirata}}, \bibinfo {author} {\bibfnamefont {T.}~\bibnamefont {Fujita}},
  \bibinfo {author} {\bibfnamefont {Y.~R.}\ \bibnamefont {Wen}}, \bibinfo
  {author} {\bibfnamefont {J.~H.}\ \bibnamefont {Schneibel}}, \bibinfo {author}
  {\bibfnamefont {C.~T.}\ \bibnamefont {Liu}}, \ and\ \bibinfo {author}
  {\bibfnamefont {M.~W.}\ \bibnamefont {Chen}},\ }\href@noop {} {\bibfield
  {journal} {\bibinfo  {journal} {Nat. Mater.}\ }\textbf {\bibinfo {volume}
  {10}},\ \bibinfo {pages} {922} (\bibinfo {year} {2011})}\BibitemShut
  {NoStop}%
\bibitem [{\citenamefont {Arsenlis}\ \emph {et~al.}(2007)\citenamefont
  {Arsenlis}, \citenamefont {Cai}, \citenamefont {Tang}, \citenamefont {Rhee},
  \citenamefont {Oppelstrup}, \citenamefont {Hommes}, \citenamefont {Pierce},\
  and\ \citenamefont {Bulatov}}]{arsenlis2007enabling}%
  \BibitemOpen
  \bibfield  {author} {\bibinfo {author} {\bibfnamefont {A.}~\bibnamefont
  {Arsenlis}}, \bibinfo {author} {\bibfnamefont {W.}~\bibnamefont {Cai}},
  \bibinfo {author} {\bibfnamefont {M.}~\bibnamefont {Tang}}, \bibinfo {author}
  {\bibfnamefont {M.}~\bibnamefont {Rhee}}, \bibinfo {author} {\bibfnamefont
  {T.}~\bibnamefont {Oppelstrup}}, \bibinfo {author} {\bibfnamefont
  {G.}~\bibnamefont {Hommes}}, \bibinfo {author} {\bibfnamefont {T.~G.}\
  \bibnamefont {Pierce}}, \ and\ \bibinfo {author} {\bibfnamefont {V.~V.}\
  \bibnamefont {Bulatov}},\ }\href@noop {} {\bibfield  {journal} {\bibinfo
  {journal} {Modell. Simul. Mater. Sci. Eng.}\ }\textbf {\bibinfo {volume}
  {15}},\ \bibinfo {pages} {553} (\bibinfo {year} {2007})}\BibitemShut
  {NoStop}%
\bibitem [{\citenamefont {Mohles}(2004)}]{mohles2004critical}%
  \BibitemOpen
  \bibfield  {author} {\bibinfo {author} {\bibfnamefont {V.}~\bibnamefont
  {Mohles}},\ }\href@noop {} {\bibfield  {journal} {\bibinfo  {journal} {Mater.
  Sci. Eng. A}\ }\textbf {\bibinfo {volume} {365}},\ \bibinfo {pages} {144}
  (\bibinfo {year} {2004})}\BibitemShut {NoStop}%
\bibitem [{\citenamefont {Monnet}\ \emph {et~al.}(2011)\citenamefont {Monnet},
  \citenamefont {Naamane},\ and\ \citenamefont {Devincre}}]{monnet2011orowan}%
  \BibitemOpen
  \bibfield  {author} {\bibinfo {author} {\bibfnamefont {G.}~\bibnamefont
  {Monnet}}, \bibinfo {author} {\bibfnamefont {S.}~\bibnamefont {Naamane}}, \
  and\ \bibinfo {author} {\bibfnamefont {B.}~\bibnamefont {Devincre}},\
  }\href@noop {} {\bibfield  {journal} {\bibinfo  {journal} {Acta Mater.}\
  }\textbf {\bibinfo {volume} {59}},\ \bibinfo {pages} {451} (\bibinfo {year}
  {2011})}\BibitemShut {NoStop}%
\bibitem [{\citenamefont {Bacon}\ \emph {et~al.}(1973)\citenamefont {Bacon},
  \citenamefont {Kocks},\ and\ \citenamefont {Scattergood}}]{bacon1973effect}%
  \BibitemOpen
  \bibfield  {author} {\bibinfo {author} {\bibfnamefont {D.~J.}\ \bibnamefont
  {Bacon}}, \bibinfo {author} {\bibfnamefont {U.~F.}\ \bibnamefont {Kocks}}, \
  and\ \bibinfo {author} {\bibfnamefont {R.~O.}\ \bibnamefont {Scattergood}},\
  }\href@noop {} {\bibfield  {journal} {\bibinfo  {journal} {Philos. Mag.}\
  }\textbf {\bibinfo {volume} {28}},\ \bibinfo {pages} {1241} (\bibinfo {year}
  {1973})}\BibitemShut {NoStop}%
\bibitem [{\citenamefont {Nordlund}\ \emph {et~al.}(1998)\citenamefont
  {Nordlund}, \citenamefont {Ghaly}, \citenamefont {Averback}, \citenamefont
  {Caturla}, \citenamefont {{Diaz de la Rubia}},\ and\ \citenamefont
  {Tarus}}]{nor97}%
  \BibitemOpen
  \bibfield  {author} {\bibinfo {author} {\bibfnamefont {K.}~\bibnamefont
  {Nordlund}}, \bibinfo {author} {\bibfnamefont {M.}~\bibnamefont {Ghaly}},
  \bibinfo {author} {\bibfnamefont {R.~S.}\ \bibnamefont {Averback}}, \bibinfo
  {author} {\bibfnamefont {M.}~\bibnamefont {Caturla}}, \bibinfo {author}
  {\bibfnamefont {T.}~\bibnamefont {{Diaz de la Rubia}}}, \ and\ \bibinfo
  {author} {\bibfnamefont {J.}~\bibnamefont {Tarus}},\ }\href@noop {}
  {\bibfield  {journal} {\bibinfo  {journal} {Phys. Rev. B}\ }\textbf {\bibinfo
  {volume} {57}},\ \bibinfo {pages} {7556} (\bibinfo {year}
  {1998})}\BibitemShut {NoStop}%
\bibitem [{\citenamefont {Ghaly}\ \emph {et~al.}(1999)\citenamefont {Ghaly},
  \citenamefont {Nordlund},\ and\ \citenamefont {Averback}}]{gha97}%
  \BibitemOpen
  \bibfield  {author} {\bibinfo {author} {\bibfnamefont {M.}~\bibnamefont
  {Ghaly}}, \bibinfo {author} {\bibfnamefont {K.}~\bibnamefont {Nordlund}}, \
  and\ \bibinfo {author} {\bibfnamefont {R.~S.}\ \bibnamefont {Averback}},\
  }\href@noop {} {\bibfield  {journal} {\bibinfo  {journal} {Phil. Mag. A}\
  }\textbf {\bibinfo {volume} {79}},\ \bibinfo {pages} {795} (\bibinfo {year}
  {1999})}\BibitemShut {NoStop}%
\bibitem [{\citenamefont {Henriksson}\ \emph {et~al.}(2013)\citenamefont
  {Henriksson}, \citenamefont {Bj\"{o}rkas},\ and\ \citenamefont
  {Nordlund}}]{hen13}%
  \BibitemOpen
  \bibfield  {author} {\bibinfo {author} {\bibfnamefont {K.~O.~E.}\
  \bibnamefont {Henriksson}}, \bibinfo {author} {\bibfnamefont
  {C.}~\bibnamefont {Bj\"{o}rkas}}, \ and\ \bibinfo {author} {\bibfnamefont
  {K.}~\bibnamefont {Nordlund}},\ }\href@noop {} {\bibfield  {journal}
  {\bibinfo  {journal} {J. Phys. Cond. Matter}\ }\textbf {\bibinfo {volume}
  {25}},\ \bibinfo {pages} {445401} (\bibinfo {year} {2013})}\BibitemShut
  {NoStop}%
\bibitem [{\citenamefont {Osetsky}\ and\ \citenamefont {Bacon}(2003)}]{ose03}%
  \BibitemOpen
  \bibfield  {author} {\bibinfo {author} {\bibfnamefont {Y.~N.}\ \bibnamefont
  {Osetsky}}\ and\ \bibinfo {author} {\bibfnamefont {D.~J.}\ \bibnamefont
  {Bacon}},\ }\href@noop {} {\bibfield  {journal} {\bibinfo  {journal} {Modell.
  Simul. Mater. Sci. Eng.}\ }\textbf {\bibinfo {volume} {11}},\ \bibinfo
  {pages} {427} (\bibinfo {year} {2003})}\BibitemShut {NoStop}%
\bibitem [{\citenamefont {Berendsen}\ \emph {et~al.}(1984)\citenamefont
  {Berendsen}, \citenamefont {Postma}, \citenamefont {van Gunsteren},
  \citenamefont {DiNola},\ and\ \citenamefont {Haak}}]{ber84}%
  \BibitemOpen
  \bibfield  {author} {\bibinfo {author} {\bibfnamefont {H.~J.~C.}\
  \bibnamefont {Berendsen}}, \bibinfo {author} {\bibfnamefont {J.~P.~M.}\
  \bibnamefont {Postma}}, \bibinfo {author} {\bibfnamefont {W.~F.}\
  \bibnamefont {van Gunsteren}}, \bibinfo {author} {\bibfnamefont
  {A.}~\bibnamefont {DiNola}}, \ and\ \bibinfo {author} {\bibfnamefont {J.~R.}\
  \bibnamefont {Haak}},\ }\href@noop {} {\bibfield  {journal} {\bibinfo
  {journal} {J. Chem. Phys}\ }\textbf {\bibinfo {volume} {81}},\ \bibinfo
  {pages} {3684} (\bibinfo {year} {1984})}\BibitemShut {NoStop}%
\bibitem [{\citenamefont {Granberg}\ \emph {et~al.}(2014)\citenamefont
  {Granberg}, \citenamefont {Terentyev}, \citenamefont {Henriksson},
  \citenamefont {Djurabekova},\ and\ \citenamefont {Nordlund}}]{gra1}%
  \BibitemOpen
  \bibfield  {author} {\bibinfo {author} {\bibfnamefont {F.}~\bibnamefont
  {Granberg}}, \bibinfo {author} {\bibfnamefont {D.}~\bibnamefont {Terentyev}},
  \bibinfo {author} {\bibfnamefont {K.~O.~E.}\ \bibnamefont {Henriksson}},
  \bibinfo {author} {\bibfnamefont {F.}~\bibnamefont {Djurabekova}}, \ and\
  \bibinfo {author} {\bibfnamefont {K.}~\bibnamefont {Nordlund}},\ }\href@noop
  {} {\bibfield  {journal} {\bibinfo  {journal} {Fusion Sci. Technol.}\
  }\textbf {\bibinfo {volume} {66}},\ \bibinfo {pages} {283} (\bibinfo {year}
  {2014})}\BibitemShut {NoStop}%
\bibitem [{\citenamefont {Granberg}\ \emph
  {et~al.}(2015{\natexlab{a}})\citenamefont {Granberg}, \citenamefont
  {Terentyev},\ and\ \citenamefont {Nordlund}}]{gra2}%
  \BibitemOpen
  \bibfield  {author} {\bibinfo {author} {\bibfnamefont {F.}~\bibnamefont
  {Granberg}}, \bibinfo {author} {\bibfnamefont {D.}~\bibnamefont {Terentyev}},
  \ and\ \bibinfo {author} {\bibfnamefont {K.}~\bibnamefont {Nordlund}},\
  }\href@noop {} {\bibfield  {journal} {\bibinfo  {journal} {J. Nucl. Mater.}\
  }\textbf {\bibinfo {volume} {460}},\ \bibinfo {pages} {23} (\bibinfo {year}
  {2015}{\natexlab{a}})}\BibitemShut {NoStop}%
\bibitem [{\citenamefont {Granberg}\ \emph
  {et~al.}(2015{\natexlab{b}})\citenamefont {Granberg}, \citenamefont
  {Terentyev},\ and\ \citenamefont {Nordlund}}]{gra3}%
  \BibitemOpen
  \bibfield  {author} {\bibinfo {author} {\bibfnamefont {F.}~\bibnamefont
  {Granberg}}, \bibinfo {author} {\bibfnamefont {D.}~\bibnamefont {Terentyev}},
  \ and\ \bibinfo {author} {\bibfnamefont {K.}~\bibnamefont {Nordlund}},\
  }\href@noop {} {\bibfield  {journal} {\bibinfo  {journal} {Nucl. Instrum.
  Methods Phys. Res., Sect. B}\ }\textbf {\bibinfo {volume} {352}},\ \bibinfo
  {pages} {77} (\bibinfo {year} {2015}{\natexlab{b}})}\BibitemShut {NoStop}%
\bibitem [{\citenamefont {Stukowski}(2010)}]{stu10}%
  \BibitemOpen
  \bibfield  {author} {\bibinfo {author} {\bibfnamefont {A.}~\bibnamefont
  {Stukowski}},\ }\href@noop {} {\bibfield  {journal} {\bibinfo  {journal}
  {Modell. Simul. Mater. Sci. Eng.}\ }\textbf {\bibinfo {volume} {18}},\
  \bibinfo {pages} {015012} (\bibinfo {year} {2010})}\BibitemShut {NoStop}%
\bibitem [{\citenamefont {Eshelby}(1957)}]{eshelby1957determination}%
  \BibitemOpen
  \bibfield  {author} {\bibinfo {author} {\bibfnamefont {J.~D.}\ \bibnamefont
  {Eshelby}},\ }in\ \href@noop {} {\emph {\bibinfo {booktitle} {Proceedings of
  the Royal Society of London A: Mathematical, Physical and Engineering
  Sciences}}},\ Vol.\ \bibinfo {volume} {241}\ (\bibinfo {organization} {The
  Royal Society},\ \bibinfo {year} {1957})\ pp.\ \bibinfo {pages}
  {376--396}\BibitemShut {NoStop}%
\bibitem [{\citenamefont {Hull}\ and\ \citenamefont
  {Bacon}(2011)}]{hull2011introduction}%
  \BibitemOpen
  \bibfield  {author} {\bibinfo {author} {\bibfnamefont {D.}~\bibnamefont
  {Hull}}\ and\ \bibinfo {author} {\bibfnamefont {D.~J.}\ \bibnamefont
  {Bacon}},\ }\href@noop {} {\emph {\bibinfo {title} {Introduction to
  dislocations}}},\ Vol.~\bibinfo {volume} {37}\ (\bibinfo  {publisher}
  {Elsevier},\ \bibinfo {year} {2011})\BibitemShut {NoStop}%
\bibitem [{\citenamefont {Murty}\ and\ \citenamefont
  {Charit}(2008)}]{murty2008structural}%
  \BibitemOpen
  \bibfield  {author} {\bibinfo {author} {\bibfnamefont {K.~L.}\ \bibnamefont
  {Murty}}\ and\ \bibinfo {author} {\bibfnamefont {I.}~\bibnamefont {Charit}},\
  }\href@noop {} {\bibfield  {journal} {\bibinfo  {journal} {J. Nucl. Mater.}\
  }\textbf {\bibinfo {volume} {383}},\ \bibinfo {pages} {189} (\bibinfo {year}
  {2008})}\BibitemShut {NoStop}%
\bibitem [{\citenamefont {Queyreau}\ \emph {et~al.}(2011)\citenamefont
  {Queyreau}, \citenamefont {Marian}, \citenamefont {Gilbert},\ and\
  \citenamefont {Wirth}}]{queyreau2011edge}%
  \BibitemOpen
  \bibfield  {author} {\bibinfo {author} {\bibfnamefont {S.}~\bibnamefont
  {Queyreau}}, \bibinfo {author} {\bibfnamefont {J.}~\bibnamefont {Marian}},
  \bibinfo {author} {\bibfnamefont {M.~R.}\ \bibnamefont {Gilbert}}, \ and\
  \bibinfo {author} {\bibfnamefont {B.~D.}\ \bibnamefont {Wirth}},\ }\href@noop
  {} {\bibfield  {journal} {\bibinfo  {journal} {Phys. Rev. B}\ }\textbf
  {\bibinfo {volume} {84}},\ \bibinfo {pages} {064106} (\bibinfo {year}
  {2011})}\BibitemShut {NoStop}%
\bibitem [{\citenamefont {Gilbert}\ \emph {et~al.}(2011)\citenamefont
  {Gilbert}, \citenamefont {Queyreau},\ and\ \citenamefont
  {Marian}}]{gilbert2011screw}%
  \BibitemOpen
  \bibfield  {author} {\bibinfo {author} {\bibfnamefont {M.~R.}\ \bibnamefont
  {Gilbert}}, \bibinfo {author} {\bibfnamefont {S.}~\bibnamefont {Queyreau}}, \
  and\ \bibinfo {author} {\bibfnamefont {J.}~\bibnamefont {Marian}},\
  }\href@noop {} {\bibfield  {journal} {\bibinfo  {journal} {Phys. Rev. B}\
  }\textbf {\bibinfo {volume} {84}},\ \bibinfo {pages} {174103} (\bibinfo
  {year} {2011})}\BibitemShut {NoStop}%
\bibitem [{\citenamefont {Ovaska}\ \emph {et~al.}(2015)\citenamefont {Ovaska},
  \citenamefont {Laurson},\ and\ \citenamefont {Alava}}]{ovaska2014collective}%
  \BibitemOpen
  \bibfield  {author} {\bibinfo {author} {\bibfnamefont {M.}~\bibnamefont
  {Ovaska}}, \bibinfo {author} {\bibfnamefont {L.}~\bibnamefont {Laurson}}, \
  and\ \bibinfo {author} {\bibfnamefont {M.~J.}\ \bibnamefont {Alava}},\
  }\href@noop {} {\bibfield  {journal} {\bibinfo  {journal} {Sci. Rep.}\
  }\textbf {\bibinfo {volume} {5}},\ \bibinfo {pages} {10580} (\bibinfo {year}
  {2015})}\BibitemShut {NoStop}%
\end{thebibliography}%

\end{document}